\documentclass[11pt,a4paper]{article}

\usepackage{graphicx}
\usepackage{amsmath}
\usepackage{amssymb}
\usepackage{amsxtra}
\usepackage{feynmp-auto}
\usepackage{simplewick}
\usepackage{jheppub}
\usepackage{subcaption}
\def\be{\begin{eqnarray}}
\def\ee{\end{eqnarray}}
\def\0{\nonumber}
\def\d{\partial}
\def\tr{{\rm tr}}
\def\trace{{\rm tr}} 
\def\del{\partial}
\def\al{\alpha}
\def\si{\sigma}
\def\la{\lambda}
\usepackage{slashed}
\usepackage{verbatim}
\usepackage{latexsym}
\usepackage{euscript}
\usepackage{soul}
\usepackage{color}
\newcommand\EC{\EuScript{C}}
\newcommand\EA{\EuScript{A}}
\newcommand\ER{\EuScript{R}}
\newcommand\ET{\EuScript{T}}
\newcommand\EL{\EuScript{L}}
\newcommand\EF{\EuScript{F}}
\newcommand\ED{\EuScript{D}}
\normalfont\large
 
\preprint{SISSA/11/2015/FISI}
 
\title{Regularization of energy-momentum tensor correlators and parity-odd
terms}

\author[a,b]{Loriano Bonora,}
\author[a,c]{Ant\^onio Duarte Pereira}
\author[a,b]{and Bruno Lima de Souza}

\affiliation[a]{International School for Advanced Studies (SISSA),\\Via Bonomea
265, 34136 Trieste, Italy}
\affiliation[b]{INFN, Sezione di Trieste, Italy}
\affiliation[c]{Universidade Federal Fluminense (UFF),\\Instituto de F\'isica,
Campus da Praia Vermelha,
Avenida General Milton Tavares de Souza s/n, 24210-346, Niter\'oi, RJ, Brasil}

\emailAdd{bonora@sissa.it}
\emailAdd{aduarte@if.uff.br}
\emailAdd{blima@sissa.it}

\abstract{We discuss the problem of regularizing correlators in conformal field theories. The 
only way to do it in coordinate space is to interpret them as distributions. Unfortunately
except for the simplest cases we do not have tabulated mathematical results. The way out
we pursue here is to go to momentum space and use Feynman diagram techniques and their
regularization methods. We focus on the energy-momentum tensor correlators and, to
gain insight, we compute and regularize $2$-point functions in $2d$ with various techniques both
in coordinate space and in momentum space, obtaining the same results. Then we
do the same for $2$-point functions in $4d$. Finally we turn to $3$-point function in $4d$,
and concentrate on the parity-odd part. We derive in particular the regularized trace and 
divergence of the energy-momentum tensor in a chiral fermion model. We discuss the problems
related to the parity-odd trace anomaly.}
 

\begin{document}
\maketitle

\section{Introduction}
\label{sec:intro}

In recent years CFT in $4d$ has been receiving an increasing attention. The
reason is well-known, it is due in part to being one of the poles in the AdS/CFT
correspondence, in part to the new applications to strongly correlated systems
and in part to the increasing interest in applying the standard model of
elementary particles to very high energy problems and its coupling to gravity.
In turn this  has spurred a lot of interest and activity in the theoretical
aspects of conformal symmetry and conformal field theories. Recent reviews on
the latter are \cite{Nakayama,Rychkov}, older references relevant to the content
of this paper are \cite{Osborn93,Osborn96}. One of the most striking recently
obtained results is the derivation of the general structures of conformal
covariant correlators and OPE's of any kind of tensor fields in coordinate
space,
\cite{Weinberg:2010fx,Costa:2011mg,Stanev1,Zhiboedov,Stanev2,Costa:2014rya,
Elkhidir:2014woa}. The analysis of $3$-point functions of conserved currents and
the energy-momentum tensor was also considered in momentum space,
\cite{Cappelli:2001pz,Coriano:2012wp,Bzowski:2013sza}. 

The above mentioned correlators in coordinate space are in general {\it
unregulated} expressions, in that they have singularities at coincident points.
For convenience we call them {\it semiclassical}. The natural way to regularize
them is provided by distribution theory. This is clear in theory, in practice it
is not so simple because, except for the simplest cases, one has to do with
formidable expressions. In the coordinate representation a rather natural
technique is provided by the so-called {\it differential regularization},
\cite{Freedman1,Freedman2,Latorre:1993xh}. However this technique does not seem
to be in general algorithmic (see below) and a good deal of guesswork is needed
in order to obtain sensible expressions.

Regularizing  correlators is not simply a procedure (legitimately) required by
mathematics. Singularities in correlators usually contain useful information.
For instance in correlators of currents or energy-momentum tensors singularities
provide information about the coupling to gauge potentials and to gravity,
respectively. This is the case of anomalies, which are a typical result of
regularization processes, though independent of them. Regularized correlators
are also necessary in the Callan-Symanzik equation, \cite{Osborn91}. In summary, regularizing
conformal correlators is the next necessary step after deriving their
(unregulated or semiclassical) expressions.

As was said above, however, the process of regularizing higher order correlators in coordinate space representation with differential regularization does not seem to be algorithmic. For definiteness we concentrate here on the $2$- and $3$-point functions of the energy-momentum tensor. We show that we have a definite rule to regularize the $2$-point correlators in coordinate space by means of differential regularization, but when we come to the $3$-point function there is a discontinuity which does not allow us to extend the rule valid for the $2$-point one. To understand the origin of the problem we resort to a model, the model of a free chiral fermion, in momentum representation. Using one-loop Feynman diagrams we can determine completely the $3$-point correlator of the 
e.m. tensor and regularize it with standard dimensional regularization techniques. The idea is to Fourier anti-transform it in order to shed light on the regularization in the coordinate representation. For two reasons we
concentrate on the parity-odd part, although the extension to the parity-even part is straightforward. The first reason is the presence of the Levi-Civita tensor which limits the number of terms to a more manageable amount, while preserving all the general features of the problem. 

The second reason is more important: the appearance of the Pontryagin density in
the trace anomaly of this model. This parity-odd anomaly has been recalculated
explicitly in \cite{BGL} after the first appearance in \cite{ChD1,ChD2}, with
different methods. If one uses Feynman diagram techniques the basic evaluation
is that of the triangle diagram. Now, it has been proved recently (this is one
of the general results mentioned above) that the parity-odd part of the
$3$-point function of the energy-momentum tensor in the coordinate
representation 
vanishes identically, \citep{Zhiboedov,Stanev1}. Therefore it would seem that there is a contradiction with
the existence of a parity-odd part in the trace of the e.m. tensor. Although
this argument is rather naive and forgetful of the subtleties of quantum field
theory, it seems to be widespread. Therefore we think it is worth clarifying it.
We show below that in fact there is no contradiction: a vanishing parity-odd
semiclassical $3$-point function of the energy-momentum tensor must in fact
coexist with a nonvanishing parity-odd part of the trace anomaly.

The paper is organized as follows. In the next three sections we thoroughly
analyse the $2d$ case. The reason is that, although the results are known, in
$2d$ many problems that will appear in higher dimensions are already present and
can be fully solved. So $2d$ is a useful playground for the rest of the paper.
In section \ref{sec:2ptFunctionIn2d} and \ref{sec:ParityOddIn2d} we consider the
problem of regularizing the $2$-point function of e.m. tensors in $2d$ using the
techniques of differential regularization (for the various techniques used, see \cite{bert,fuji,BvN}). In section \ref{sec:FeynmanDiagrams2d} we analyze the $2$-point function of the
e.m. tensor using Feynman diagrams techniques. In section \ref{sec:2ptFunction4d}
we compute the $2$-point function of e.m. tensors in $4d$ both using
differential regularization and Feynman diagrams. In section
\ref{sec:3ptCorrelator} we review a general {\it no-go} argument concerning
parity-odd contributions in the $3$-point function of e.m. tensors, we
explicitly compute the parity-odd part of the correlator of three e.m. tensors
in the chiral fermion model in $4d$ in coordinate representation and show that
it identically vanishes. We repeat the last computation using Feynman diagrams
and regularize it, and show how it gives rise to the parity-odd trace anomaly.
We show that irreducible Lorentz components of the correlators, in particular
those containing the trace and the traceless part of the e.m. tensor, must be
regularized separately. We also discuss the connection of the anomaly with the
e.m. conservation. We show that in general regularization breaks covariance and
counterterms must be subtracted in order to recover it. In section
\ref{sec:UglyDuckling} we discuss the prejudices on the existence of the
Pontryagin anomaly.

To complete this introduction we present general formulas for the trace and
divergence of the e.m. tensor. The problem of regularizing the e.m. correlators
is strictly connected with (and  clarified by) coupling the system to gravity.

\subsection{General formulas for the trace and divergence of the e.m. tensor}
\label{ssec:genform}

In general let us couple the energy-momentum tensor of a theory to a classical
external source
$j_{\mu\nu}$. The partition function in terms of $j$ is
\be 
Z[j^{\mu\nu}]&=& \langle 0|\mathcal{T}\{\,e^{\frac{i}{2}\int dx
T_{\mu\nu}(x)j^{\mu\nu}(x)}\}|0\rangle=
e^{-iW[j_{\mu\nu}]}\label{Zg}\\
&=& \sum_{n=0}^\infty \frac {i^n}{2^n n!} \int  \prod_{i=1}^n dx_i \, \,
j^{\mu_i\nu_i}(x_i)\, 
\langle 0|\mathcal{T}\{T_{\mu_1\nu_1}(x_1)\ldots
T_{\mu_n\nu_n}(x_n)\}|0\rangle,\0
\ee
where the symbol $\mathcal{T}$ denotes a time-ordered product. The generating functional of connected Green functions is
\be 
W[j^{\mu\nu}]= \sum_{n=1}^\infty \frac {i^{n+1}}{2^n n!} \int \prod_{i=1}^n
dx_i\, 
j^{\mu_i\nu_i}(x_i)\, \langle 0|\mathcal{T}\{T_{\mu_1\nu_1}(x_1)\cdots
T_{\mu_n\nu_n}(x_n)\}|0\rangle_c .\label{Wg}
\ee
We will denote the full one-loop e.m. tensor by 
\be 
\langle\!\langle T_{\mu\nu}(x)\rangle\!\rangle = 
  \left. 2\frac {\delta W[j]}{\delta
j^{\mu\nu}(x)}\right|_{j^{\mu\nu}=h^{\mu\nu}}, \label{qem}
\ee
where $h_{\mu\nu}$ is the fluctuation, $g_{\mu\nu}=
\eta_{\mu\nu}+h_{\mu\nu}+\dots$ and $g^{\mu\nu}= \eta^{\mu\nu}-
h^{\mu\nu}+\dots$, with respect to the flat metric $\eta_{\mu\nu}$ \footnote{The factor 
$\frac 1{2^n}$ in (\ref{Zg}) is motivated by the fact that when we expand the action 
$$S[\eta+h]= S[\eta] + \int d^dx \frac {\delta S}{\delta g^{\mu\nu}}\Big{\vert}_{g=\eta} h^{\mu\nu}+\cdots,$$ 
the factor $ \frac {\delta S}{\delta g^{\mu\nu}}\Big{\vert}_{g=\eta}= \frac 12 T_{\mu\nu}$. 
Another consequence of this fact will be that the presence of vertices with one graviton in 
Feynman diagrams will correspond to insertions of the operator $\frac{1}{2}T_{\mu\nu}$ 
in correlation functions.}. The fluctuation $h_{\mu\nu}$ is the field attached 
to the external legs in the Feynman diagrams approach.  We can reconstruct the 
full one-loop e.m. tensor as a function of $h_{\mu\nu}$ by means of the formula
\be
\begin{aligned}\label{reconstruction}
\langle\!\langle T_{\mu\nu}(x)\rangle\!\rangle=\frac 1{n!} \sum_{n=1}^\infty
\int  dx_1\ldots & \int dx_n  h^{\mu_1\nu_1}(x_1) \ldots h^{\mu_n\nu_n}(x_n)\\
& \times \left. \frac{\delta}{\delta h^{\mu_1\nu_1}(x_1)}\ldots
\frac{\delta}{\delta h^{\mu_n\nu_n}(x_n)}\langle\!\langle
T_{\mu\nu}(x)\rangle\!\rangle\right|_{h=0}.
\end{aligned}\ee
For instance, to first and second order in $h$ the trace is given by  
\be 
\left. \frac{\delta}{\delta h^{\lambda\rho}(y)} \langle\!\langle
T_\mu^\mu(x)\rangle\!\rangle\right|_{h=0} =2i\,\langle 0|\mathcal{T}\{
T_\mu^\mu(x)
T_{\lambda\rho}(y)\} |0\rangle \label{1sttrace}
\ee
and
\be
\left. \frac {\delta}{\delta h^{\lambda\rho}(y)}\frac{\delta}{\delta
h^{\alpha\beta}(z)}\langle\!\langle
T_\mu^\mu(x)\rangle\!\rangle\right|_{h=0}&=&-2i
\left(\delta^{(4)}(x-y)+\delta^{(4)}(x-z)\right)\langle
0|\mathcal{T}\{T_{\lambda\rho}(y)T_{\alpha\beta}(z)\}|0\rangle \0 \\
&&+ 2\langle
0|\mathcal{T}\{T_\mu^\mu(x)T_{\lambda\rho}(y)T_{\alpha\beta}(z)\}|0\rangle ,
\label{2ndtrace}
\ee
and the divergence by 
\be
\left.\frac {\delta}{\delta h^{\lambda\rho}(y)} \langle\!\langle\nabla^\mu
T_{\mu\nu}(x)\rangle\!\rangle\right|_{h=0} =-2i\,\langle 0|\mathcal{T}\{\del^\mu
T_{\mu\nu} (x)T_{\lambda\rho}(y)\}|0\rangle\label{1stdivergence}
\ee
and
\be \label{2nddivergence}
&&\left. \frac {\delta}{\delta h^{\lambda\rho}(y)}\frac {\delta}{\delta
h^{\alpha\beta}(z)}\langle\!\langle \nabla^\mu
T_{\mu\nu}(x)\rangle\!\rangle\right|_{h=0}= \0\\
&& i\left\{ \frac{\del}{\del x^{(\alpha}}\left[ \delta(x-z)\langle
0|\mathcal{T}\{T_{\beta)\nu}(x) T_{\lambda\rho}(y)\}|0\rangle\right] + 
\frac{\del}{\del x^{(\lambda}}  \left[ \delta(x-y)\langle
0|\mathcal{T}\{T_{\rho)\nu}(x) T_{\al\beta}(z)\}|0\rangle\right] \right.\0\\
&&+2\,\frac{\del}{\del x_\tau}\delta(x-z)\eta_{\al\beta} \langle
0|\mathcal{T}\{T_{\tau\nu}(x)T_{\lambda\rho}(y)\}|0\rangle + 2\,\frac
{\del}{\del x_\tau}\delta(x-y)\eta_{\lambda\rho} \langle
0|\mathcal{T}\{T_{\tau\nu}(x)T_{\al\beta}(z)\}|0\rangle\0\\
&&+ \left. 2\,\frac{\del}{\del{x^\nu}}
\delta(x-z)\langle 0|\mathcal{T}\{T_{\lambda\rho}(y)T_{\alpha\beta}(x)\}|0\rangle + 2\,\frac{\del}{\del{x^\nu}}
\delta(x-y)\langle 0|\mathcal{T}\{T_{\lambda\rho}(x)T_{\alpha\beta}(z)\}|0\rangle\right\}\0\\
&&+ 2\,\langle 0|\mathcal{T}\{\del^\mu T_{\mu\nu}
(x)T_{\lambda\rho}(y)T_{\alpha\beta}(z)\}|0\rangle,
\ee
respectively,
where the delta functions are $4$-dimensional and the round brackets indicate
symmetrization. These formulas are obtained understanding that gravity is
minimally coupled and that the background is flat. If there is a nontrivial
background metric, say $g^{(0)}_{\mu\nu}$, then we must insert $\sqrt{g^{(0)}}$
in the integral in the exponent of (\ref{Zg}) and, for instance,
(\ref{2ndtrace}) would be replaced by
\be 
&&\left. \frac 1{\sqrt{g^{(0)}(y)}}\frac {\delta}{\delta h^{\lambda\rho}(y)}\,
\frac 1{\sqrt{g^{(0)}(z)} } \frac {\delta}{\delta
h^{\alpha\beta}(z)}\langle\!\langle
T_\mu^\mu(x)\rangle\!\rangle\right|_{h=0}\label{2ndtraceg0}\\
&&=-2i \left(\frac {\delta^{(4)}(x-y)}{\sqrt{g^{(0)}(y)}}+\frac
{\delta^{(4)}(x-z)}{\sqrt{g^{(0)}(z)}}\right)\langle
0|\mathcal{T}\{T_{\lambda\rho}(y)T_{\alpha\beta}(z)\}|0\rangle  + \langle
0|\mathcal{T}\{T_\mu^\mu(x)T_{\lambda\rho}(y)T_{\alpha\beta}(z)\}|0\rangle \0
\ee
and (\ref{2nddivergence}) by a much more complicated formula.

\section{$2$-point function of e.m. tensors in $2d$ and trace anomaly}
\label{sec:2ptFunctionIn2d}

In this section we regularize the $2$-point function of energy-momentum tensors
in $2d$ using the techniques 
of differential regularization and we derive the very well-known $2d$ trace
anomaly. The ambiguities implicit in the regularization procedure allow us to
make manifest the interplay between diffeomorphism and trace anomalies.

Let us consider the $2$-point function $\left\langle
T_{\mu\nu}\left(x\right)T_{\rho\sigma}\left(0\right)\right\rangle $. 
This $2$-point function in $2d$ (i.e. the semiclassical $2$-point function) is
very well-known and is given by\footnote{One way of deriving this expression is by using the embedding formalism, see \cite{Weinberg:2010fx}, for example.}
\begin{equation}
\left\langle T_{\mu\nu}\left(x\right)T_{\rho\sigma}\left(0\right)\right\rangle
=\frac{c/2}{x^{4}}\left(I_{\mu\rho}\left(x\right)I_{\nu\sigma}\left(x\right)+I_{
\nu\rho}\left(x\right)I_{\mu\sigma}\left(x\right)-\eta_{\mu\nu}\eta_{\rho\sigma}
\right)\label{eq:Goal}
\end{equation}
where
\begin{equation}
I_{\mu\nu}\left(x\right)=\eta_{\mu\nu}-2\frac{x_{\mu}x_{\nu}}{x^{2}}\label{
eq:Imunu}
\end{equation}
and $c$ is the central charge of the theory. For $x\neq0$ this $2$-point
function satisfies the Ward identities
\begin{align}
\partial^{\mu}\left\langle
T_{\mu\nu}\left(x\right)T_{\rho\sigma}\left(0\right)\right\rangle  &
=0,\label{eq:Conservation}\\
\left\langle
T_{\mu}^{\mu}\left(x\right)T_{\rho\sigma}\left(0\right)\right\rangle  &
=0.\label{eq:Tracelessness}
\end{align}
The result (\ref{eq:Goal}) is obtained using the symmetry properties of the indices, 
dimensional analysis and eqs. (\ref{eq:Conservation}) and (\ref{eq:Tracelessness}).

The $2$-point function written above are UV singular for $x\rightarrow0$, hence
this divergence has to be dealt with 
for the correlator to be well-defined everywhere. In this context the most
convenient way to regularize this object is 
with the technique of \emph{differential regularization.} The recipe of
differential regularization is: given a 
function $f\left(x\right)$ that needs to be regularized, find the most general
function $F\left(x\right)$ such 
that $\mathcal{D}F\left(x\right)=f\left(x\right)$, where $\mathcal{D}$ is some
differential operator, and such that 
the Fourier transform of $\mathcal{D}F\left(x\right)$ is well-defined
(alternatively $\mathcal{D}F\left(x\right)$ has 
integrable singularities).

In our case we have two guiding principles: the Ward identities and dimensional analysis. 
Differential regularization tells that our $2$-point function should be some differential 
operator applied to a function, i.e.
\begin{equation}
\left\langle T_{\mu\nu}\left(x\right)T_{\rho\sigma}\left(0\right)\right\rangle
=\mathcal{D}_{\mu\nu\rho\sigma}\left(f\left(x\right)\right),\label{eq:Ansatz2ptsFct2d}
\end{equation}
while conservation requires that the differential operator
$\mathcal{D}_{\mu\nu\rho\sigma}$
be transverse, i.e.
\begin{equation}
\partial^{\mu}\mathcal{D}_{\mu\nu\rho\sigma}=\dots=\partial^{\sigma}\mathcal{D}_
{\mu\nu\rho\sigma}=0.
\end{equation}
The most general transverse operator with four derivatives, symmetric in
$\mu$,$\nu$ and in $\rho$,$\sigma$ that one can write is
\begin{equation}
\mathcal{D}_{\mu\nu\rho\sigma}=\alpha\mathcal{D}_{\mu\nu\rho\sigma}^{
\left(1\right)}+\beta\mathcal{D}_{\mu\nu\rho\sigma}^{\left(2\right)},
\end{equation}
where
\begin{align}
\mathcal{D}_{\mu\nu\rho\sigma}^{\left(1\right)} &=
\partial_{\mu}\partial_{\nu}\partial_{\rho}\partial_{\sigma}-\left(\eta_{\mu\nu
}\partial_{\rho}\partial_{\sigma}+\eta_{\rho\sigma}\partial_{\mu}\partial_{\nu}
\right)\Box+\eta_{\mu\nu}\eta_{\rho\sigma}\Box\Box,\\ \nonumber
\mathcal{D}_{\mu\nu\rho\sigma}^{\left(2\right)}&=
\partial_{\mu}\partial_{\nu}\partial_{\rho}\partial_{\sigma}-\frac{1}{2}
\left(\eta_{\mu\rho}\partial_{\nu}\partial_{\sigma}+\eta_{\nu\rho}\partial_{\mu}
\partial_{\sigma}+\eta_{\mu\sigma}\partial_{\nu}\partial_{\rho}+\eta_{\nu\sigma}
\partial_{\mu}\partial_{\rho}\right)\Box\\ 
&\quad+\frac{1}{2}\left(\eta_{\mu\rho}\eta_{
\nu\sigma}+\eta_{\nu\rho}\eta_{\mu\sigma}\right)\Box\Box.
\end{align}
One important fact about these differential operators is that they may not be
traceless. Indeed, by taking the trace we find
\begin{equation}
\eta^{\mu\nu}\mathcal{D}_{\mu\nu\rho\sigma}^{\left(1\right)}=\eta^{\mu\nu}
\mathcal{D}_{\mu\nu\rho\sigma}^{\left(2\right)}=-\left(\partial_{\rho}\partial_{
\sigma}-\eta_{\rho\sigma}\Box\right)\Box.
\end{equation}
Dimensional analysis tells us that the function $f\left(x\right)$ in
\eqref{eq:Ansatz2ptsFct2d} can be at most a function of
$\log\mu^{2}x^{2}$ since the lhs of \eqref{eq:Ansatz2ptsFct2d} scales like
$1/x^{4}$ and this scaling is already saturated by 
the differential operator with four derivatives. Notice that we have introduced
an arbitrary mass scale $\mu$ to make 
the argument of the $\log$ dimensionless. Let us write the most general ansatz
for \eqref{eq:Ansatz2ptsFct2d}: 
\begin{eqnarray}
\left\langle T_{\mu\nu}\left(x\right)T_{\rho\sigma}\left(0\right)\right\rangle 
& = &
\phantom{+}\mathcal{D}_{\mu\nu\rho\sigma}^{\left(1\right)}\left[\alpha_{1}
\log\mu^{2}x^{2}+\alpha_{2}
\left(\log\mu^{2}x^{2}\right)^{2}+\cdots\right]\nonumber \\
&&+\mathcal{D}_{\mu\nu\rho\sigma}^{\left(2\right)}\left[\beta_{1}\log\mu^{2}x^{2
}+\beta_{2}
\left(\log\mu^{2}x^{2}\right)^{2}+\cdots\right].\label{eq:GeneralAnsatz}
\end{eqnarray}
Now our task is to fix the coefficients $\alpha_{i}$ and $\beta_{j}$ for
\eqref{eq:GeneralAnsatz} to match 
\eqref{eq:Goal} for $x\neq 0$. As it turns out we only need terms up to
$\log^{2}$ (otherwise one cannot avoid 
logarithmic terms for $x\neq 0$)  The matching gives us
\[
\alpha_{1}=-\frac{c}{24}-\beta_{1},\quad\alpha_{2}=-\beta_{2}=-\frac{c}{96}, 
\]
thus
\begin{equation}
\left\langle T_{\mu\nu}\left(x\right)T_{\rho\sigma}\left(0\right)\right\rangle
=-\frac{c}{24}\mathcal{D}_{\mu\nu\rho\sigma}^{\left(1\right)}
\left(\log\mu^{2}x^{2}\right)-\frac{c}{96}\left(\mathcal{D}_{
\mu\nu\rho\sigma}^{\left(1\right)}-\mathcal{D}_{\mu\nu\rho\sigma}^{
\left(2\right)}\right)\left(\log\mu^{2}x^{2}\right)^{2}.
\label{eq:Solution}
\end{equation}
Notice that $\beta_{1}$ is absent in the final result. Indeed, the term with
coefficient $\beta_{1}$ is 
\be
-\left(\mathcal{D}_{\mu\nu\rho\sigma}^{\left(1\right)}-\mathcal{D}_{
\mu\nu\rho\sigma}^{\left(2\right)}\right)\left(\log\mu^{2}x^{2}\right)\label{
CZterm}
\ee
and this term identically vanishes in $2d$. If we take the trace of
\eqref{eq:Solution} we find that
\[
\left\langle
T_{\mu}^{\mu}\left(x\right)T_{\rho\sigma}\left(0\right)\right\rangle
=-\frac{c}{48}\eta^{\mu\nu}\mathcal{D}_{\mu\nu\rho\sigma}^{\left(1\right)}
\left(\log\mu^{2}x^{2}\right)=\frac{c}{48}\left(\partial_{\rho}
\partial_{\sigma}-\eta_{\rho\sigma}\Box\right)\Box\log\mu^{2}x^{2}.
\]
These terms have support only at $x=0$, for in $2d$ the d'Alembertian of a
$\log$ is a delta function, more precisely
\begin{equation}
\Box\log\mu^{2}x^{2}=4\pi\delta^{2}\left(x\right).
\end{equation}
Therefore we find the anomalous Ward identity
\begin{equation}\label{2ptTrace2d}
\left\langle
T_{\mu}^{\mu}\left(x\right)T_{\rho\sigma}\left(y\right)\right\rangle
=c\frac{\pi}{12}\left(\partial_{\rho}\partial_{\sigma}-\eta_{\rho\sigma}
\Box\right)\delta^{2}\left(x-y\right),
\end{equation}
If we consider our theory in the presence of a background metric $g$ which is a
perturbation of flat spacetime, 
i.e. $g_{\rho\sigma}(y)=\eta_{\rho\sigma}+h_{\rho\sigma}(y)+\cdots$, eq.
(\ref{2ptTrace2d}) gives rise 
to the lowest contribution to the `full one-loop' trace of the e.m. tensor,
namely
\be
\langle\!\langle
T_{\mu}^{\mu}\rangle\!\rangle=c\frac{\pi}{12}(\del_\mu\del_\nu-\eta_{\mu\nu}
\Box)h^{\mu\nu},
\ee
which coincides with the lowest contribution of the expansion in $h$ of the
Ricci scalar, i.e.
\be
R=(\del_\mu\del_\nu-\eta_{\mu\nu}\Box)h^{\mu\nu}+\mathcal{O}(h^2).
\ee
Covariance requires that the higher order corrections in $h$ to the `full
one-loop' trace of the e.m. tensor in the presence 
of a background metric $g$ to be such that we recover the covariant expression
\begin{equation}
\langle\!\langle T_{\mu}^{\mu}\rangle\!\rangle
=c\frac{\pi}{12}R.\label{2dtraceanomalt}
\end{equation}
For a free chiral fermion $c=1/4\pi^2$, vide section \ref{sec:FeynmanDiagrams2d}
or appendix \ref{sec:2dchiralmodel}. We are authorized to use the covariant
expression (\ref{2dtraceanomalt}) because the energy-momentum tensor is
conserved 
(there are no diffemorphism anomalies).

Using the above results it is easy to verify the Callan-Symanzik equation for
the $2$-point function (\ref{eq:Solution}). 
The Callan-Symanzik differential operator reduces to the logarithmic derivative
with respect to $\mu$, because both beta functions 
and anomalous dimensions vanish in the case we are considering. We get
\be 
\mu \frac {\del}{\del \mu} \left\langle
T_{\mu\nu}\left(x\right)T_{\rho\sigma}\left(0\right)\right\rangle \sim
\left(\mathcal{D}_{\mu\nu\rho\sigma}^{\left(1\right)}-\mathcal{D}_{
\mu\nu\rho\sigma}^{\left(2\right)}\right)\left(\log\mu^{2}x^{2}\right)=0.\label{
CZ}
\ee
We see that requiring that the regularized correlator satisfies conservation at
$x=0$ implies the appearance of a trace anomaly. 
However this is not the end of the story, since there are ambiguities in the
regularization process we have so far disregarded.

\subsection{Ambiguities}

The ambiguity arises from the fact that we can add to (\ref{eq:Solution}) terms
that have support only in $x=0$. 
The most general modification of the parity-even part that would affect only its
expression for $x=0$ is given by
\begin{eqnarray}
A_{\mu\nu\rho\sigma} & = &
\phantom{+}A\left(\eta_{\mu\nu}\partial_{\rho}\partial_{\sigma}+\eta_{\rho\sigma
}\partial_{\mu}\partial_{\nu}\right)\Box\log\mu^{2}x^{2}\nonumber \\
&  &
+B\left(\eta_{\mu\rho}\partial_{\nu}\partial_{\sigma}+\eta_{\nu\rho}\partial_{
\mu}\partial_{\sigma}+\eta_{\mu\sigma}\partial_{\nu}\partial_{\rho}+\eta_{
\nu\sigma}\partial_{\mu}\partial_{\rho}\right)\Box\log\mu^{2}x^{2}\nonumber \\
&  &
+C\left(\eta_{\mu\rho}\eta_{\nu\sigma}+\eta_{\nu\rho}\eta_{\mu\sigma}
\right)\Box\Box\log\mu^{2}x^{2}\nonumber \\
 &  &
+D\eta_{\mu\nu}\eta_{\rho\sigma}\Box\Box\log\mu^{2}x^{2}.\label{eq:EvenAmbig}
\end{eqnarray}
We remark that this term is in general neither conserved nor traceless
\begin{eqnarray}
\del^\mu A_{\mu\nu\rho\sigma}&=& 4 \pi\left((A+2B)\del_\nu\del_\rho\del_\sigma
+(A+D) \eta_{\rho\sigma} \del_\nu\Box \right.\0\\
&&+\left. (B+C) \left(\eta_{\rho\nu}\del_\sigma
\Box+\eta_{\sigma\nu} \del_{\rho}
\Box\right)\right)\delta^{(2)}(x)\label{ambdiff}\\
A^{\mu}_{\mu\rho\sigma}&=&4\pi \left((2A+4B)\del_\rho\del_\sigma+
(A+2C+2D)\eta_{\rho\sigma} \Box\right) \delta^{(2)}(x)\label{ambtrace}
\end{eqnarray}

We notice that by imposing (\ref{ambdiff}) to vanish imply that also
(\ref{ambtrace}) will vanish. We may wonder whether using this ambiguity we can
cancel the trace anomaly. This can certainly be done by choosing
$2A+4B=-A-2C-2D$ and adjusting  the overall coefficient. But this operation
gives rise to a diffeomorphism anomaly. Its form is far from appealing and not
particularly illuminating, so we do not write it down (see however
\cite{BBP1,BBP2}). In  other words the anomaly (\ref{2dtraceanomalt}) is a
non-trivial cocycle of the overall 
symmetry diffeomorphisms plus Weyl transformations. As was discussed in
\cite{BBP1,BBP2} it may take different forms, either as a pure diffeomorphism
anomaly or a pure trace anomaly. In general both components may be nonvanishing.
It is obvious that, in practice, it is more useful to preserve diffeomorphism
invariance, so that the cocycle takes the form  (\ref{2dtraceanomalt}). 

\section{Parity-odd terms in $2d$}
\label{sec:ParityOddIn2d}

In this section we compute all possible semiclassical parity-odd terms in the
$2$-point function of the energy-momentum tensor in $2d$. 
We follow three methods, the first two are general while the third is based on a
specific model. Needless to say all methods give the 
same results up to ambiguities.

\subsection{Using symmetries}

The first method is very simple-minded, it consists in writing the most general
expression $ {\cal T}^{\rm odd}_{\mu\nu\rho\sigma}(x)$ linear 
in the antisymmetric tensor $\epsilon_{\alpha\beta}$ with the right dimensions
which is symmetric and traceless in $\mu,\nu$ and 
$\rho,\sigma$ separately, is symmetric in the exchange $(\mu,\nu)\leftrightarrow
(\rho,\sigma)$, and is conserved. The calculation 
is tedious but straightforward. The result is as follows. Let us define
\be 
T_{\mu\nu\rho\sigma}= \frac 1{x^4}\left(I_{\mu\rho}(x)I_{\nu\sigma}(x)+
I_{\mu\sigma}(x)I_{\nu\rho}(x)-\eta_{\mu\nu}\eta_{\rho\sigma}\right),\label{Smnrs}
\ee
and
\be
{\cal T}^{\rm odd}_{\mu\nu\rho\sigma}(x)=\frac {\mathfrak e}4
\left(\epsilon_{\mu\lambda} T^\lambda{}_{\nu\rho\sigma}\left(x\right)+
\epsilon_{\nu\lambda}T_\mu{}^\lambda{}_{\rho\sigma}\left(x\right)+
\epsilon_{\rho\lambda}T_{\mu\nu}{}^\lambda{}_{\sigma}\left(x\right) +
\epsilon_{\sigma\lambda}T_{\mu\nu\rho}{}^\lambda\left(x\right)\right).\label{Todd}
\ee
where ${\mathfrak e}$ is an undetermined constant. We assume (\ref{Todd}) to
represent 
$\langle T_{\mu\nu}(x) T_{\rho\sigma}(0)\rangle_{\rm odd}$. It satisfies all the
desired properties (it is traceless and conserved). 
In order to make sure that it is conformal covariant, we have to check that it
is chirally split. To this end we introduce the 
light-cone coordinates $x_\pm= x^0\pm x^1$. It is not hard to verify that
\be 
\langle T_{++}(x) T_{--}(0)\rangle _{\rm odd}=0.\label{chirality}
\ee

\subsection{The embedding formalism}

The second method is the embedding formalism
\cite{Weinberg:2010fx,Costa:2011mg}, which consists in using the fact that
conformal covariance in $d$ 
dimensions can be linearly realized in $d+2$. After constructing a covariant
expression in $d+2$ one projects to $d$ dimensional 
Minkowski space. In particular for $d=2$ the method works as follows. We write
the most general parity-odd contribution to the 
$2$-point function of a symmetric $2$-tensor in $4d$ which, in addition, is
transverse:
\begin{equation}
\left\langle T_{AB}\left(X\right)T_{CD}\left(Y\right)\right\rangle
_{\text{odd}}=\frac{1}{\left(X\cdot
Y\right)^{2}}\left[\epsilon_{AICJ}\frac{X^{I}Y^{J}}{X\cdot
Y}\left(\eta_{BD}-\frac{X_{D}Y_{B}}{X\cdot Y}\right)+A\leftrightarrow
B\right]+C\leftrightarrow D.\label{eq:POddNullCone}
\end{equation}
This term is symmetric on $A$, $B$ and $C$, $D$ and is transverse with respect
to $X_{A}$, $X_{B}$, $Y_{C}$ and $Y_{D}$. Our next step is to project this
quantity to $2d$. The projected correlator is given by
\begin{equation}
\left\langle T_{\mu\nu}\left(x\right)T_{\rho\sigma}\left(y\right)\right\rangle
_{\text{odd}}=\frac{\partial X^{A}}{\partial x^{\mu}}\frac{\partial
X^{B}}{\partial x^{\nu}}\frac{\partial Y^{C}}{\partial y^{\rho}}\frac{\partial
Y^{D}}{\partial y^{\sigma}}\left\langle
T_{AB}\left(X\right)T_{CD}\left(Y\right)\right\rangle
_{\text{odd}}.\label{eq:Projection}
\end{equation}
We recall that
\begin{equation}
\frac{\partial X^{A}}{\partial
x^{\mu}}=\delta_{-}^{A}2x_{\mu}+\delta_{\mu}^{A}\equiv\left(0,2x_{\mu},\delta_{
\mu}^{a}\right),\quad A=+,-,a.
\end{equation}
The contractions with the $\epsilon$-tensor give rise to a determinant, namely
\begin{equation}
\epsilon_{AICJ}\frac{\partial X^{A}}{\partial x^{\mu}}X^{I}\frac{\partial
Y^{C}}{\partial y^{\rho}}Y^{J}\equiv\left|\begin{array}{cccc}
0 & 1 & 0 & 1\\
2x_{\mu} & x^{2} & 2y_{\rho} & y^{2}\\
\delta_{\mu}^{a} & x^{i} & \delta_{\rho}^{c} & y^{j}
\end{array}\right|.\label{eq:Determinant}
\end{equation}
The translational invariance of the problem allows us to rewrite it in the form
\begin{equation}
\left|\begin{array}{cccc}
0 & 1 & 0 & 1\\
2\left(x-y\right)_{\mu} & \left(x-y\right)^{2} & 0 & 0\\
\delta_{\mu}^{a} & \left(x-y\right)^{i} & \delta_{\rho}^{c} & 0
\end{array}\right|=-\left|\begin{array}{ccc}
2\left(x-y\right)_{\mu} & \left(x-y\right)^{2} & 0\\
\delta_{\mu}^{a} & \left(x-y\right)^{i} & \delta_{\rho}^{c}
\end{array}\right|.\label{eq:TransInvAssump}
\end{equation}
For convenience, let us relabel $x-y\rightarrow x$. This determinant is
straightforward to compute and it gives us
\begin{equation}
-\left|\begin{array}{ccc}
2x_{\mu} & x^{2} & 0\\
\delta_{\mu}^{a} & x^{i} & \delta_{\rho}^{c}
\end{array}\right|=-\left(2x_{\mu}\left|\begin{array}{cc}
x^{i} & \delta_{\rho}^{c}\end{array}\right|-x^{2}\left|\begin{array}{cc}
\delta_{\mu}^{a} &
\delta_{\rho}^{c}\end{array}\right|\right)=-\left(2x_{\mu}\epsilon_{\alpha\rho}
x^{\alpha}-x^{2}\epsilon_{\mu\rho}\right).
\end{equation}
Thus, the projected correlator is given by
\begin{equation}
\left\langle T_{\mu\nu}\left(x\right)T_{\rho\sigma}\left(0\right)\right\rangle
_{\text{odd}}=\frac{e}{x^{4}}\left[\epsilon_{\alpha\rho}\left(\delta_{\mu}^{
\alpha}-2\frac{x_{\mu}x^{\alpha}}{x^{2}}\right)\left(\eta_{\nu\sigma}-2\frac{x_{
\nu}x_{\sigma}}{x^{2}}\right)+\mu\leftrightarrow\nu\right]
+\rho\leftrightarrow\sigma.
\end{equation}
In terms of $I_{\mu\nu}\left(x\right)$ we have
\be\begin{aligned}
\left\langle
T_{\mu\nu}\left(x\right)T_{\rho\sigma}\left(0\right)\right\rangle_{\text{odd}}
&=\frac{e}{x^{4}}\left[\epsilon_{\alpha\rho}\left(I_{\mu}^{\alpha}\left(x\right)
I_{\nu\sigma}\left(x\right)+I_{\nu}^{\alpha}\left(x\right)I_{\mu\sigma}
\left(x\right)\right)\right.\\
&\phantom{=\;}\left. \quad + \epsilon_{\alpha\sigma} \left(
I_{\mu}^{\alpha}\left(x\right) I_{\nu\rho}\left(x\right) +
I_{\nu}^{\alpha}\left(x\right) I_{\mu\rho}\left(x\right) \right) \right].
\label{eq:WithTransInv}
\end{aligned}\ee
This correlator satisfies both tracelessness and conservation, as it can be
verified by a direct computation, but it is not symmetric under the exchange of
$\mu,\nu$ with $\rho,\sigma$. Thus, our final expression is
\eqref{eq:WithTransInv} symmetrized in
$\left(\mu,\nu\right)\leftrightarrow\left(\rho,\sigma\right)$:
\be\begin{aligned}
\left\langle
T_{\mu\nu}\left(x\right)T_{\rho\sigma}\left(0\right)\right\rangle_{\text{odd}} &
= \frac{e}{x^{4}} \left[ \epsilon_{\alpha\mu}
\left(I_{\rho}^{\alpha}\left(x\right) I_{\nu\sigma}\left(x\right) +
I_{\sigma}^{\alpha}\left(x\right) I_{\nu\rho}\left(x\right)\right)\right. \\ 
&\phantom{=\quad\;}  + \epsilon_{\alpha\nu}
\left(I_{\rho}^{\alpha}\left(x\right) I_{\mu\sigma}\left(x\right) +
I_{\sigma}^{\alpha}\left(x\right) I_{\mu\rho}\left(x\right) \right)\\
&\phantom{=\quad\;} + \epsilon_{\alpha\rho} \left(
I_{\mu}^{\alpha}\left(x\right) I_{\nu\sigma}\left(x\right) +
I_{\nu}^{\alpha}\left(x\right) I_{\mu\sigma}\left(x\right) \right)\\
&\phantom{=\quad\;}\left. + \epsilon_{\alpha\sigma}
\left(I_{\mu}^{\alpha}\left(x\right) I_{\nu\rho}\left(x\right) +
I_{\nu}^{\alpha}\left(x\right) I_{\mu\rho}\left(x\right) \right)
\right].\label{eq:WithTransInvSym}
\end{aligned}\ee
From \eqref{eq:WithTransInvSym} we notice a tensorial structure very similar to
the parity-even part of the $2$-point function of $T_{\mu\nu}$, namely
\begin{equation}
T_{\mu\nu\rho\sigma}\left(x\right)=\frac{1}{x^{4}}\left(I_{\mu\rho}
\left(x\right)I_{\nu\sigma}\left(x\right)+I_{\nu\rho}\left(x\right)I_
{\mu\sigma}\left(x\right)-\eta_{\mu\nu}\eta_{\rho\sigma}\right)\label{
eq:EvenConstructor}
\end{equation}
and it turns out that we may write \eqref{eq:WithTransInvSym} in terms of the
partity-even part, i.e.
\begin{equation}
\left\langle T_{\mu\nu}\left(x\right)T_{\rho\sigma}\left(0\right)\right\rangle
_{\text{odd}}=\frac{e}{2}\left(\epsilon_{\alpha\mu}T_{\phantom{\alpha}
\nu\rho\sigma}^{\alpha}\left(x\right)+\epsilon_{\alpha\nu}T_{\mu\phantom{\alpha}
\rho\sigma}^{
\phantom{\mu}\alpha}\left(x\right)+\epsilon_{\alpha\rho}T_{\mu\nu\phantom{\alpha
}\sigma}^{
\phantom{\mu\nu}\alpha}\left(x\right)+\epsilon_{\alpha\rho}T_{\mu\nu\rho\phantom
{\alpha}}^{
\phantom{\mu\nu\rho}\alpha}\left(x\right)\right).\label{eq:NullConeExpression}
\end{equation}
This result looks different from (\ref{Todd}) but it is not hard to show that,
for $x\neq 0$, they are proportional: ${\mathfrak e}= \frac 34 e$

Still another method to derive the same result is to use a free fermion model.
This is deferred to appendix \ref{sec:2dchiralmodel}.

\subsection{Differential regularization of the parity-odd part}

The task of regularizing the parity-odd terms is very much simplified by the
fact that we are able to write them in terms of the parity-even part, see
(\ref{eq:NullConeExpression}). We can therefore use the same regularization as
in section \ref{sec:2ptFunctionIn2d}. Let us start by the regularization that
preserves diffeomorphisms for the parity-even part, eq. (\ref{eq:Solution}):
\begin{equation}
T_{\mu\nu\rho\sigma}\left(x\right)=-\frac{1}{12}\mathcal{D}_{\mu\nu\rho\sigma}^{
\left(1\right)
}\left(\log\mu^{2}x^{2}\right)-\frac{1}{48}\left(\mathcal{D}_{
\mu\nu\rho\sigma}^{\left(1\right)}-\mathcal{D}_{\mu\nu\rho\sigma}^{
\left(2\right)}\right)\left(\log\mu^{2}x^{2}\right)^{2}.\label{eq:ConservedReg}
\end{equation}
Regularizing \eqref{eq:NullConeExpression} with \eqref{eq:ConservedReg} leads to
a trace anomaly
\begin{equation}
\left\langle
T_{\mu}^{\mu}\left(x\right)T_{\rho\sigma}\left(0\right)\right\rangle
_{\textrm{odd}}=\frac{\pi
e}{24}\left(\epsilon_{\rho\alpha}\partial^{\alpha}\partial_{\sigma}+\epsilon_{
\sigma\alpha}\partial^{\alpha}\partial_{\rho}\right)\delta^{2}\left(x\right),
\label{eq:TraceOdd1}
\end{equation}
and a diffeomorphism anomaly
\begin{equation}
\partial^{\mu}\left\langle
T_{\mu\nu}\left(x\right)T_{\rho\sigma}\left(0\right)\right\rangle
_{\textrm{odd}}=\frac{\pi
e}{24}\epsilon_{\nu\alpha}\partial^{\alpha}\left(\eta_{\rho\sigma}\Box-\partial_
{\rho}\partial_{\sigma}\right)\delta^{2}\left(x\right).\label{eq:DivOdd1}
\end{equation}
In the presence of a background metric $g$ the anomalous Ward-Identities
(\ref{eq:TraceOdd1}) and (\ref{eq:DivOdd1}) 
give rise to the following `full one-loop' functions
\be  
\langle\!\langle T_\mu^\mu(x)\rangle\!\rangle & = & \frac{\pi e}{24}
 \epsilon^{\lambda\alpha}\del_\alpha \left(g^{\rho\sigma}\del_\lambda
g_{\rho\sigma} + g^{\rho\sigma}\del_\rho g_{\lambda\sigma}\right), 
\label{traceanom}\\
\langle\!\langle \nabla^\mu  T_{\mu\nu}(x)\rangle\!\rangle &=& \frac{\pi e}{24}
\epsilon_{\nu\alpha} \del^\alpha R.
\label{diffanomcov}
\ee
The second is the well-known covariant form of the diffeomorphism anomaly. The
consistent form of the same anomaly is 
\be 
\langle\!\langle\nabla^\mu  T_{\mu\nu}(x)\rangle\!\rangle \sim
\epsilon^{\mu\rho} \del_\mu\del_\alpha \Gamma_{\rho\nu}^\alpha. 
\label{diffanomconsist}
\ee
We remark however that in $2d$ the two forms (\ref{diffanomcov}) and
(\ref{diffanomconsist}) collapse to the same form to the lowest order, 
since
\be 
2 \epsilon_{\mu\nu} \del^\mu\left( \del_\alpha\del_\beta- \eta_{\alpha\beta}
\Box\right)=
\epsilon_{\mu\alpha}\left(\del^\mu\del_\nu \del_\beta -\eta_{\nu\beta}
\del^\mu\Box + (\alpha \leftrightarrow \beta)\right)\0
\ee
We see that, in any case, the diffeomorphism anomaly is accompanied by the a
trace anomaly.

\subsection{Ambiguities in the parity-odd part}

We know that the regularization used above is not the ultimate one, because
there are ambiguities. They entail a modification of the parity-odd part given
by
\begin{equation}
A_{\mu\nu\rho\sigma}^{\text{odd}}=\epsilon_{\alpha\mu}A_{\phantom{\alpha}
\nu\rho\sigma}^{\alpha}+\epsilon_{\alpha\nu}A_{\mu\phantom{\alpha}\rho\sigma}^{
\phantom{\mu}\alpha}+\epsilon_{\alpha\rho}A_{\mu\nu\phantom{\alpha}\sigma}^{
\phantom{\mu\nu}\alpha}+\epsilon_{\alpha\rho}A_{\mu\nu\rho\phantom{\alpha}}^{
\phantom{\mu\nu\rho}\alpha},
\end{equation}
where the RHS is written in terms of \eqref{eq:EvenAmbig}, which explicitly is
\be\begin{aligned}
A_{\mu\nu\rho\sigma}^{\text{odd}} & = \phantom{+} A \left[ \eta_{\mu\nu} \left(
\epsilon_{\rho\alpha} \partial^{\alpha} \partial_{\sigma} +
\epsilon_{\sigma\alpha} \partial^{\alpha} \partial_{\rho}\right)
+\eta_{\rho\sigma} \left( \epsilon_{\mu\alpha} \partial^{\alpha} \partial_{\nu}
+ \epsilon_{\nu\alpha} \partial^{\alpha} \partial_{\mu} \right) \right] \Box
\log\mu^{2}x^{2}\\
&\phantom{=} + B \left[ \epsilon_{\mu\alpha} \left( \eta_{\nu\rho}
\partial^{\alpha} \partial_{\sigma} + \eta_{\nu\sigma} \partial^{\alpha}
\partial_{\rho} \right) + \epsilon_{\nu\alpha} \left( \eta_{\mu\rho}
\partial^{\alpha} \partial_{\sigma} + \eta_{\mu\sigma} \partial^{\alpha}
\partial_{\rho} \right) \right. \\
&\phantom{= B +} \left. + \epsilon_{\rho\alpha} \left( \eta_{\sigma\mu}
\partial^{\alpha} \partial_{\nu} + \eta_{\sigma\nu} \partial^{\alpha}
\partial_{\mu} \right) + \epsilon_{\sigma\alpha} \left( \eta_{\rho\mu}
\partial^{\alpha} \partial_{\nu} + \eta_{\rho\nu} \partial^{\alpha}
\partial_{\mu} \right) \right] \Box\log\mu^{2}x^{2}.\label{eq:OddAmbig}
\end{aligned}\ee
The trace and the divergence of \eqref{eq:OddAmbig} are given by:

\begin{equation}
\eta^{\mu\nu}A_{\mu\nu\rho\sigma}=8\pi\left(A+2B\right)\left(\epsilon_{
\rho\alpha}\partial^{\alpha}\partial_{\sigma}+\epsilon_{\sigma\alpha}\partial^{
\alpha}\partial_{\rho}\right)\delta^{2}\left(x\right),\label{eq:OddAmbigTrace}
\end{equation}
\be \begin{aligned}
\partial^{\mu}A_{\mu\nu\rho\sigma} & =
\phantom{+}4\pi\left(B\eta_{\nu\rho}\Box+\left(A+B\right)\partial_{\nu}\partial_
{\rho}\right)\epsilon_{\sigma\alpha}\partial^{\alpha}\delta^{2}
\left(x\right) \\
&\phantom{=}
+4\pi\left(B\eta_{\nu\sigma}\Box+\left(A+B\right)\partial_{\nu}\partial_{\sigma}
\right)\epsilon_{\rho\alpha}\partial^{\alpha}\delta^{2}\left(x\right)
\\
&\phantom{=}
+4\pi\left(A\eta_{\rho\sigma}\Box+2B\partial_{\rho}\partial_{\sigma}
\right)\epsilon_{\nu\alpha}\partial^{\alpha}\delta^{2}\left(x\right).\label{
eq:OddAmbigDiv}
\end{aligned}\ee
Using these ambiguities we can recast the expressions \eqref{eq:TraceOdd1} and
\eqref{eq:DivOdd1} in the form
\begin{equation}
\left\langle
T_{\mu}^{\mu}\left(x\right)T_{\rho\sigma}\left(0\right)\right\rangle
_{\textrm{odd}}=\left(8\pi\left(A+2B\right)+\frac{\pi
e}{24}\right)\left(\epsilon_{\rho\alpha}\partial^{\alpha}\partial_{\sigma}
+\epsilon_{\sigma\alpha}\partial^{\alpha}\partial_{\rho}\right)\delta^{2}
\left(x\right),\label{eq:TraceOdd1+Amb}
\end{equation}
\be \begin{aligned}
\partial^{\mu}\left\langle
T_{\mu\nu}\left(x\right)T_{\rho\sigma}\left(0\right)\right\rangle
_{\textrm{odd}} & =
4\pi\left(B\eta_{\nu\rho}\Box+\left(A+B\right)\partial_{\nu}\partial_{\rho}
\right)\epsilon_{\sigma\alpha}\partial^{\alpha}\delta^{2}\left(x\right)
\\
&\phantom{=}
+4\pi\left(B\eta_{\nu\sigma}\Box+\left(A+B\right)\partial_{\nu}\partial_{\sigma}
\right)\epsilon_{\rho\alpha}\partial^{\alpha}\delta^{2}\left(x\right)
\\
&\phantom{=} +\epsilon_{\nu\alpha}\partial^{\alpha}\left(\left(4\pi A+\frac{\pi
e}{24}\right)\eta_{\rho\sigma}\Box+\left(8\pi B-\frac{\pi
e}{24}\right)\partial_{\rho}\partial_{\sigma}\right)\delta^{2}
\left(x\right).\label{eq:DivOdd1+Amb}
\end{aligned}\ee
If we impose that \eqref{eq:TraceOdd1+Amb} is zero we find
\be \label{eq:TracelessRegOption}
A=-\frac{e}{192}-2B,
\ee
which implies that \eqref{eq:DivOdd1+Amb} takes the form
\be \begin{aligned} \label{eq:FinalTracelessOption}
\partial^{\mu}\left\langle
T_{\mu\nu}\left(x\right)T_{\rho\sigma}\left(0\right)\right\rangle
_{\textrm{odd}}  & = 
4\pi\left[B\eta_{\nu\rho}\Box-\left(\frac{e}{192}+B\right)\partial_{\nu}
\partial_{\rho}\right]\epsilon_{\sigma\alpha}\partial^{\alpha}\delta^{2}
\left(x\right)\\
&\phantom{=}
+4\pi\left[B\eta_{\nu\sigma}\Box-\left(\frac{e}{192}+B\right)\partial_{\nu}
\partial_{\sigma}\right]\epsilon_{\rho\alpha}\partial^{\alpha}\delta^{2}
\left(x\right)\\
&\phantom{=} -\epsilon_{\nu\alpha}\partial^{\alpha}\left[\left(\frac{\pi
e}{48}+8\pi B \right) \eta_{\rho\sigma} \Box - \left(8\pi B-\frac{\pi
e}{24}\right)\partial_{\rho}\partial_{\sigma}\right]\delta^{2}\left(x\right).
\end{aligned}\ee
The choice (\ref{eq:TracelessRegOption}) allows us to eliminate the trace
anomaly (\ref{traceanom}) but by doing so the diffeo anomaly becomes
(\ref{eq:FinalTracelessOption}), which will not imply a covariant expression for
$\langle\!\langle T_{\mu\nu} \rangle\!\rangle$ for any choice of $B$. Thus, the
most general regularization that one can write is given by the equations
\eqref{eq:TraceOdd1+Amb} and \eqref{eq:DivOdd1+Amb}. An important point of
\eqref{eq:DivOdd1+Amb} is that there is no choice of $A$ and $B$ for which it is
zero, hence inevitably we will have a diffeomorphism anomaly, unless the overall
factor $e=0$, which depends of course on the specific model.

\section{The Feynman diagrams method in $2d$}
\label{sec:FeynmanDiagrams2d}

\begin{figure}[h]
\centering
\begin{fmffile}{BubbleDiagram}
	\begin{fmfgraph*}(100,60)
		\fmfleft{i} \fmflabel{$\mu,\nu$}{i}
		\fmfright{f} \fmflabel{$\lambda,\rho$}{f}
		\fmf{photon,label=$k$}{i,v1}
		\fmf{photon,label=$k$}{v2,f}
		\fmf{fermion,left,tension=.4,label=$p$}{v1,v2}
		\fmf{fermion,left,tension=.4,label=$p-k$}{v2,v1}
		\fmfdot{v1,v2}
	\end{fmfgraph*}
\end{fmffile}
\caption{The relevant Feynman diagram for the computation.}
\label{img:BubbleD}
\end{figure}
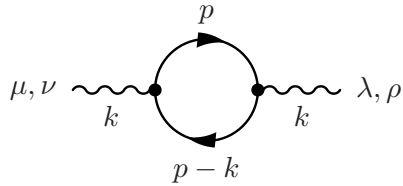

It is interesting and instructive to derive the results above using Feynman
diagrams. 
There is only one non-trivial contribution that comes from the bubble diagram
with one incoming and one outgoing line with momentum $k$ and an internal
momentum $p$ (see figure \ref{img:BubbleD}). The pertinent Feynman rule is
\begin{equation}
\parbox{20mm}{\begin{fmffile}{Vffg}
	\begin{fmfgraph*}(40,40)
		\fmfleft{i}\fmfright{f1,f2}
		\fmflabel{$\mu,\nu$}{i}
		\fmf{photon}{i,v}
		\fmf{fermion,label=$p'$,l.side=right}{v,f1}\
		\fmf{fermion,label=$p$,l.side=right}{f2,v}
	\end{fmfgraph*}
\end{fmffile}} \!\!\!\!\!\!\!\!\!\!  = 
\dfrac{i}{8}\left[
\left(p+p'\right)_\mu\gamma_\nu+\left(p+p'\right)_\nu\gamma_\mu\right]\frac{
1+\gamma_*}{2}.
\label{FRule1}
\end{equation}

The relevant $2$-point function is\footnote{The factor of 4 in (\ref{TmnTlr2d}) is 
produced by the fact that the vertex (\ref{FRule1}) corresponds to the insertion 
of $\frac 12 T_{\mu\nu}$, not simply $T_{\mu\nu}$, 
in the correlator, as explained in the footnote in (\ref{ssec:genform}).}
\be
\langle T_{\mu\nu}(x) T_{\la\rho}(y)\rangle =4 \int
\frac{d^2k}{(2\pi)^2}e^{-ik(x-y)} \ET_{\mu\nu\la\rho}(k)\label{TmnTlr2d}
\ee
with
\be
\ET_{\mu\nu\la\rho}(k)=-\frac 1{64} \int  \frac{d^2k}{(2\pi)^2} {\rm tr}
\left(\frac 1 {\slashed{p}} (2p-k)_\mu\gamma_\nu \frac
1{\slashed{p}-\slashed{k}} (2p-k)_\la \gamma_\rho \frac {1+\gamma_\ast}2\right)+
\left\{ \begin{array}{c}\mu\leftrightarrow \nu \\ \la \leftrightarrow
\rho\end{array} \right\}\label{T(k)2d}
\ee
Taking the trace and regularizing by introducing extra components of the
momentum running around the loop, $p\to p+\ell$
($\ell=\ell_2,\ldots,\ell_{\delta+2}$), we get
\be \begin{aligned}
{\ET}^\mu_{\phantom{\mu}\mu \lambda\rho }(k) & = -\frac 1 {32} \int \frac
{d^2p}{(2\pi)^2} \int \frac {d^{\delta}\ell}{(2\pi)^{\delta}} \,\tr\left(
\frac{\slashed{p}+\slashed{\ell}}{p^2-\ell^2} \, \left(2\slashed{p}+2
\slashed{\ell} -\slashed{q}\right) \right.\\
&\phantom{=-\frac{1}{8}\hspace{3.7cm}}\left. 
\frac {\slashed{p}+\slashed{\ell}-\slashed{k}}{(p-k)^2-\ell^2}(2p-k)_\lambda
\gamma_\rho
\frac {1+\gamma_\ast}2 \right)\label{T42d}
\end{aligned}\ee
and the symmetrization $\la\leftrightarrow \rho$ is understood from now on.
Introducing, as usual, a Feynman parametrization of the integral in (\ref{T42d})
and 
using the results in appendix (\ref{sec:regformulas}) one finally gets for the
even part 
\be
\left({\ET}_{\rm{even}}\right)^\mu_{\phantom{\mu}\mu \lambda\rho }(k )=\frac
1{192\pi} \left(\eta_{\lambda\rho}k^2 +k_\lambda k_\rho\right),\label{ETmumueven}
\ee
which corresponds to the trace anomaly
\be 
\langle\!\langle T_{\mu}^\mu\rangle\!\rangle = -\frac{1}{48\pi} \left(\Box
h+\del_\la \del_\rho h^{\la\rho}\right) +
\mathcal{O}\left(h^2\right).\label{Aeomega2d}
\ee
For the odd part we get instead
\be
\left(\ET_{\rm{odd}}\right)^\mu_{\phantom{\mu}\mu \lambda\rho }(k )=-\frac
1{192\pi} \left(\epsilon ^\sigma{}_\rho k_\sigma k_\lambda+
(\epsilon ^\sigma{}_\lambda k_\sigma k_\rho\right),\label{ETmumuodd}
\ee
which corresponds to the trace anomaly
\be 
\langle\!\langle T_{\mu}^\mu\rangle\!\rangle = \frac 1{24\pi}\epsilon
^\sigma{}_\rho\, \del_\sigma \del_\la h^{\la\rho}+
\mathcal{O}\left(h^2\right).\label{Aoomega2d}
\ee
The trace anomaly (\ref{Aeomega2d}) is not the expected covariant one. The only
possible explanation is that our regularization has broken 
diffeomorphism invariance. In order to check that we have to compute the
divergence of the energy-momentum tensor with the same method. 
The relevant Feynman diagram contribution is (after regularization)
\be \begin{aligned}
\ED_{\nu\la\rho}(k) &= - \frac 1{64} \int \frac {d^2p}{(2\pi)^2}\int \frac
{d^{\delta}\ell}{(2\pi)^{\delta}} \\
&\phantom{=-\frac{1}{16}}\tr\left( \frac
{\slashed{p}+\slashed{\ell}}{p^2-\ell^2}(2p-k)_\mu k^\mu\; \gamma_\nu 
\frac {\slashed{p}+\slashed{\ell}-\slashed{k}}{(p-k)^2-\ell^2}(2p-k)_\lambda
\gamma_\rho\,\frac {1+\gamma_\ast}2  \right.\\
&\phantom{=-\frac{1}{16}\;} +  \left.\frac
{\slashed{p}+\slashed{\ell}}{p^2-\ell^2}(2p-k)_\nu\, \slashed{k}\,
\frac {\slashed{p}+\slashed{\ell}-\slashed{k}}{(p-k)^2-\ell^2}(2p-k)_\lambda
\gamma_\rho\,\frac {1+\gamma_\ast}2  \right).\label{EDnlr}
\end{aligned}\ee
Explicit evaluation gives for the even part
\be 
\left(\ED_{\rm{even}}\right)_{\nu\la\rho}(k)= - \frac 1{96\pi} \eta_{\la\rho}
k_\nu k^2,\label{EDenlr}
\ee
which corresponds to the diffeomorphism anomaly
\be 
\nabla^{\mu}\langle\!\langle T_{\mu\nu}\rangle\!\rangle = \frac 1{12\pi} \xi^\nu
\del_\nu \square h +\mathcal{O}\left(h^2\right).
\label{Aexi2d}
\ee 
For the odd part we get instead
\be 
\left(\ED_{\rm{odd}}\right)_{\nu\la\rho}(k)= - \frac
1{192\pi}\,k^\sigma\left(\epsilon_{\sigma\rho}\eta_{\nu\la} -
\epsilon_{\nu\rho}  k_\la  \right)k^2 +\{\la \leftrightarrow
\rho\},\label{EDonlr}
\ee
which corresponds to the anomaly
\be 
\nabla^{\mu}\langle\!\langle T_{\mu\nu}\rangle\!\rangle = -\frac
1{96\pi}\epsilon^{\sigma\rho} \left(\del_\sigma \del_\la\del_\nu
h^\la_\rho-\del_\sigma \square h_{\rho\nu} \right).\label{Aoxi2d}
\ee

Using the lowest order Weyl transformation
\be 
\delta_\omega h_{\mu\nu}= 2\omega\, \eta_{\mu\nu},\label{BRSTWeyl}
\ee
and diffeo transformation
\be 
\delta_\xi h_{\mu\nu}= \del_\mu\xi_\nu+\del_\nu\xi_\mu, \label{BRSTDiffeo}
\ee
it is easy to prove that the consistency relations
\be
 \delta_\omega \EA_\omega = 0, \quad\quad \delta_\xi \EA_\omega+\delta_\xi
\EA_\omega = 0,\quad\quad \delta_\xi \EA_\xi = 0, \label{CCeven}
 \ee
hold, where
\be
{\EA}_\omega =- \int d^2 x \; \omega \langle\!\langle
T_{\mu}^{\mu}\rangle\!\rangle,\quad\text{and}\quad{\EA}_\xi = \int d^2 x \; \xi^\nu
\nabla^{\mu}\langle\!\langle T_{\mu\nu}\rangle\!\rangle.
\ee
For the even part ${\EA}^{(e)}$ it is possible to add a counterterm to the
action and restore covariance. The couterterm is
\be 
\EC =-\frac 1{96\pi}\! \int d^2x\,h \square h.\label{evencount2d}
\ee
After this operation the divergence of the e.m. tensor vanishes and the trace
anomaly becomes
\be 
 {\EA}^{(e)}_\omega \rightarrow {\EA}^{(e)}_\omega +\delta_\omega \EC= \frac
1{48\pi} \int\! d^2x\, \omega \left( \del_\la \del_\rho h^{\la\rho} -\square
h\right),
\label{Atrace2d}
\ee
which is the expected one (see above). 

Similarly the parity-odd anomalies (\ref{ETmumuodd}) and (\ref{Aoxi2d}) satisfy the
consistency relations (\ref{CCeven}). One can add an odd counterterm to
eliminate 
the odd trace anomaly but this is definitely a less interesting operation.

The results obtained in this section are well-known. The methods we have used to
derive them teach us important
lessons. The first concerns dimensional regularization. If not explicitly stated
it is often understood in the literature that
dimensional regularization of Feynman diagrams leads to covariant results. We
have seen explicitly that this is not true, and 
a reconstruction of covariance with counterterms is inevitable. 
In view of the discussion on $3$-points correlator of the e.m. tensor in section
\ref{ssec:3ptsodd} we notice that
the piece of (\ref{T42d})
\be
\Delta{\ET}^\mu_{\mu \lambda\rho }(k)\!
=\! -\frac 1 {8} \int\! \frac {d^2p}{(2\pi)^2}
\int\! \frac {d^{\delta}\ell}{(2\pi)^{\delta}} \,\tr\left( \frac
{\slashed{p}+\slashed{\ell}}{p^2-\ell^2} \, 2 \slashed{\ell} 
\frac {\slashed{p}+\slashed{\ell}-\slashed{k}}{(p-k)^2-\ell^2}(2p-k)_\lambda
\gamma_\rho
\frac {1+\gamma_\ast}2 \right)\label{DeltaT42d}
\ee
contributes in an essential way to both even and odd anomalies. Without this
piece the result of the calculation would be inconsistent.
It marks the difference between first regularizing and then taking the trace of
the e.m. tensor or first taking the trace and then 
regularizing.
From the above it is obvious that the second procedure is the correct one. In
other words every irreducible Lorentz component of tensors must 
be regularized separately. 
This is the second important lesson. We will return to this point also in the
final section.

\section{$2$-point correlator of e.m. tensors in $4d$}
\label{sec:2ptFunction4d}

In this section we are going to discuss the $2$-point correlator of the e.m.
tensors in $4d$. The expression in coordinate representation is well-known. We
would like here to regularize it with the differential regularization method,
and, later on, compare it with the expression obtained in momentum space with
Feynman diagram techniques.

\subsection{Differential regularization of the correlator}

The unregulated $2$-point function of e.m. tensors in arbitrary dimension $d$ in
coordinate representation is given by

\begin{equation}
\left\langle T_{\mu\nu}\left(x\right)T_{\rho\sigma}\left(0\right)\right\rangle 
=\frac{c/2}{x^{2d}}\left(I_{\mu\rho}\left(x\right)I_{\nu\sigma}\left(0\right)
+I_{\nu\rho}\left(x\right)I_{\mu\sigma}\left(0\right)-\frac{2}{d}\eta_{\mu\nu}
\eta_{\rho\sigma}\right)\label{eq:2ptArbD}
\end{equation}
where
\begin{equation}
I_{\mu\nu}\left(x\right)=\eta_{\mu\nu}-2\frac{x_{\mu}x_{\nu}}{x^{2}}.
\end{equation}
As before, it can be regularized by writing down a differential operator which,
acting on an integrable function, generates it for $x\neq0$. One possibility for
$d\geq3$ is the following\footnote{Notice that for $d>4$, the function
$1/x^{2d-4}$ is indeed integrable, while we have a function which is $\log$
divergent for $d=4$ and
linearly divergent for $d=3$ and in both cases we need a regularization. In the
spirit of differential regularization, we may use the following identities
\begin{align*}
d=3:\quad & \frac{1}{x^{2}}=\frac{1}{2}\Box\log\mu^{2}x^{2},\\
d=4:\quad & \frac{1}{x^{4}}=-\frac{1}{4}\Box\frac{\log\mu^{2}x^{2}}{x^{2}},
\end{align*}
where $\log\mu^{2}x^{2}$ and $\left(\log\mu^{2}x^{2}\right)/x^{2}$
are integrable functions in the respective dimension.}
\begin{eqnarray}
\left\langle T_{\mu\nu}\left(x\right)T_{\rho\sigma}\left(0\right)\right\rangle 
& = &
-\frac{c/2}{2\left(d-2\right)^{2}d\left(d^{2}-1\right)}\mathcal{D}_{
\mu\nu\rho\sigma}^{\left(1\right)}\left(\frac{1}{x^{2d-4}}\right)\nonumber \\
 &  &
+\frac{c/2}{2\left(d-2\right)^{2}d\left(d+1\right)}\mathcal{D}_{\mu\nu\rho\sigma
}^{\left(2\right)}\left(\frac{1}{x^{2d-4}}\right),\label{eq:2ptArbitraryDReg}
\end{eqnarray}
where
\begin{align}
\mathcal{D}_{\mu\nu\rho\sigma}^{\left(1\right)} &
=\partial_{\mu}\partial_{\nu}\partial_{\rho}\partial_{\sigma}-\left(\eta_{\mu\nu
}\partial_{\rho}\partial_{\sigma}+\eta_{\rho\sigma}\partial_{\mu}\partial_{\nu}
\right)\Box+\eta_{\mu\nu}\eta_{\rho\sigma}\Box\Box,\label{eq:op1}\\
\mathcal{D}_{\mu\nu\rho\sigma}^{\left(2\right)} &
=\partial_{\mu}\partial_{\nu}\partial_{\rho}\partial_{\sigma}-\frac{1}{2}
\left(\eta_{\mu\rho}\partial_{\nu}\partial_{\sigma}+\eta_{\nu\rho}\partial_{\mu}
\partial_{\sigma}+\eta_{\mu\sigma}\partial_{\nu}\partial_{\rho}+\eta_{\nu\sigma}
\partial_{\mu}\partial_{\rho}\right)\Box\nonumber \\
 &
\phantom{=}+\frac{1}{2}\left(\eta_{\mu\rho}\eta_{\nu\sigma}+\eta_{\nu\rho}\eta_{
\mu\sigma}\right)\Box\Box.\label{eq:op2}
\end{align}
Both these operators are conserved but not traceless:
\begin{align}
\eta^{\mu\nu}\mathcal{D}_{\mu\nu\rho\sigma}^{\left(1\right)} &
=-\left(d-1\right)\left(\partial_{\rho}\partial_{\sigma}-\eta_{\rho\sigma}
\Box\right)\Box,\\
\eta^{\mu\nu}\mathcal{D}_{\mu\nu\rho\sigma}^{\left(2\right)} &
=-\left(\partial_{\rho}\partial_{\sigma}-\eta_{\rho\sigma}\Box\right)\Box,
\end{align}
nonetheless \eqref{eq:2ptArbitraryDReg} is both conserved and traceless. The
expression \eqref{eq:2ptArbitraryDReg} coincides with \eqref{eq:2ptArbD} for
$x\neq0$, it is conserved and traceless.

There are, as usual, ambiguities in the definitions of the operators
\eqref{eq:op1} and \eqref{eq:op2} for $x=0$. Particularly, in $d=4$ we may
consider the most general modification that one could add to
the expression \eqref{eq:2ptArbitraryDReg}, namely 
\begin{align}
\mathcal{A}_{\mu\nu\rho\sigma} & =
\left[A\partial_{\mu}\partial_{\nu}\partial_{\rho}\partial_{\sigma}\Box+
B\left(\eta_{\mu\rho}\partial_{\nu}\partial_{\sigma}+\eta_{\nu\rho}\partial_{\mu
}\partial_{\sigma} + \eta_{\mu\sigma}\partial_{\nu}\partial_{\rho}+
\eta_{\nu\sigma}\partial_{\mu}\partial_{\rho}\right)\Box^{2}\right.\0 \\
 &
\phantom{=}\left.+C\left(\eta_{\mu\nu}\partial_{\rho}\partial_{\sigma}+\eta_{
\rho\sigma}\partial_{\mu}\partial_{\nu}\right)\Box^{2}
+D\left(\eta_{\mu\rho}\eta_{\nu\sigma}+\eta_{\nu\rho}\eta_{\mu\sigma}
\right)\Box^{3} + E\eta_{\mu\nu}\eta_{\rho\sigma}\Box^3\right]\frac{1}{x^{2}}.
\label{eq:ambi}
\end{align}
Conservation of $\mathcal{A}$ requires
\be
C=-A+2D,\quad D=-B,\quad E=A+2B.
\ee
With these conditions the trace of $\mathcal{A}$ is
\be
\mathcal{A_{\phantom{\mu}\mu\rho\sigma}^{\mu}}=-4\pi^{2}
\left(3A+4B\right)\left(\eta_{\rho\sigma}\Box-\partial_{\rho}\partial_{\sigma}
\right)\Box\delta\left(x\right).
\ee
This corresponds to the trivial anomaly $\Box R$, which can be subtracted away
by adding a local Weyl invariant counterterm to the action. The existence of a
definition of our differential operators which do not imply in the existence of
this anomaly reflects the fact that it is a trivial anomaly.

\subsection{$2$-point correlator with Feynman diagrams}

The computation is very similar to the one in $2d$. Again, the only diagram that
contributes is the one of figure \ref{img:BubbleD} and we have\footnote{For the factor of $4$ in (\ref{TmnTlr}),
 see the footnote in section \ref{sec:FeynmanDiagrams2d}.}
\be
\langle T_{\mu\nu}(x) T_{\la\rho}(y)\rangle = 4 \int
\frac{d^4k}{(2\pi)^4}e^{-ik(x-y)} \tilde
\ET_{\mu\nu\la\rho}(k)\label{TmnTlr}
\ee
where
\be
 \tilde\ET_{\mu\nu\la\rho}(k)=\! -\frac 1{64} \int\!\!  \frac{d^4p}{(2\pi)^4}
{\rm tr}
\left(\frac 1 {\slashed{p}} (2p-k)_\mu\gamma_\nu \frac
1{\slashed{p}-\slashed{k}} (2p-k)_\la \gamma_\rho \frac {1+\gamma_5}2\right)+
\left\{ \begin{array}{c}\mu\leftrightarrow \nu \\ \la \leftrightarrow
\rho\end{array} \right\}\label{T(k)}
\ee
        
To evaluate it we use dimensional regularization. After introducing the Feynman
parameter $x$ and shifting $p$ as follows: 
$p\to p-(1-x)k$, (\ref{T(k)}) writes\footnote{We use the mostly minus signature
for the metric.}
\be
 \tilde \ET_{\mu\nu\la\rho}(k)&=&-\frac 1{32} \int_0^1dx \int 
\frac{d^4p}{(2\pi)^4} \int  \frac{d^\delta \ell}{(2\pi)^\delta}
\frac{(2p+(1-2x)k)_\mu
(2p+(1-2x)k)_\la}{(p^2+x(1-x)k^2-\ell^2)^2}\label{T(k)1}\\
&&\times\left[(p+(1-x)k)^\si(p-xk)^\tau
\left(\eta_{\si\nu}\eta_{\tau\rho}-\eta_{\si\tau}\eta_{\nu\rho}+\eta_{\si\rho}
\eta_{\nu\tau}-i\epsilon_{\si\nu\tau\rho}\right)-\ell^2 \eta_{\nu\rho}\right]
\0
\ee
After the integrations (first $\ell$, then $p$, then $x$) one finds\footnote{To
do integration properly we have to
Wick rotate the momenta and, after integration rotate them back to the
Lorentzian signature. We understand this here.}
\be
\tilde \ET_{\mu\nu\la\rho}(k)=\tilde \ED_{\mu\nu\la\rho}(k)+\tilde
\EF_{\mu\nu\la\rho}(k)+\tilde \EL_{\mu\nu\la\rho}(k)\label{T(k)2}
\ee
where
\be \begin{aligned}
\tilde \ED_{\mu\nu\la\rho}(k)=& -\frac i{32(4\pi)^2} \frac 1 {15\delta}\left[ 
8k_\mu k_\nu k_\la k_\rho+4k^2 \left(k_\mu k_\nu \eta_{\la\rho}+k_\la k_\rho
\eta_{\mu\nu}\right)\right.\\
&-6 k^2 \left(k_\mu k_\la \eta_{\nu\rho} +k_\nu k_\la \eta_{\mu\rho}+ k_\mu
k_\rho \eta_{\nu\la}+k_\nu k_\rho \eta_{\mu\la}\right)\\
&-\left. 4k^4 \eta_{\mu\nu}\eta_{\la\rho}+6k^4\left(\eta_{\mu\la}\eta_{\nu\rho}
+\eta_{\mu\rho}\eta_{\nu\la}\right) \right]
\label{D(k)}
\end{aligned}\ee
which is divergent for $\delta\to 0$, but conserved and traceless,
\be \begin{aligned}
\tilde \EL_{\mu\nu\la\rho}(k)=&  -\frac i{32(4\pi)^2} \frac {\log k^2} {30}\left[
8k_\mu k_\nu k_\la k_\rho+4k^2 \left(k_\mu k_\nu \eta_{\la\rho}+k_\la k_\rho
\eta_{\mu\nu}\right)\right.\\
&-6 k^2 \left(k_\mu k_\la \eta_{\nu\rho} +k_\nu k_\la \eta_{\mu\rho}+ k_\mu
k_\rho \eta_{\nu\la}+k_\nu k_\rho \eta_{\mu\la}\right)\\
&-\left. 4k^4 \eta_{\mu\nu}\eta_{\la\rho}+6k^4\left(\eta_{\mu\la}\eta_{\nu\rho}
+\eta_{\mu\rho}\eta_{\nu\la}\right)\right]
\label{L(k)}
\end{aligned}\ee
which is also conserved and traceless, and
\be \begin{aligned}
\tilde \EF_{\mu\nu\la\rho}(k)=& -\frac i{32(4\pi)^2} \frac 1 {30}\left[
8\left(\gamma -\log 4\pi +\frac {31}{450} \right) k_\mu k_\nu k_\la
k_\rho\right.\\
&+2\left( 1- \gamma +  \log 4\pi+\frac{31}{150}\right) k^2 \left(k_\mu k_\la
\eta_{\nu\rho}
+k_\nu k_\la \eta_{\mu\rho}+ k_\mu k_\rho \eta_{\nu\la}+k_\nu k_\rho
\eta_{\mu\la}\right)\\
&+ k^4  \left( \frac {10}3- 4\gamma+ 4\, \log 4\pi-
\frac{47}{225}\right)\eta_{\mu\nu}\eta_{\la\rho}\\
&- k^4\left( \frac {17}3- 6\gamma + 6\, \log
4\pi\right)\left(\eta_{\mu\la}\eta_{\nu\rho}
+\eta_{\mu\rho}\eta_{\nu\la}\right)\\
&\left.-k^2 \left( 4- 4\gamma+ 4 \,\log 4\pi+ \frac{47}{450}\right)
\left(k_\mu k_\nu \eta_{\la\rho}+k_\la k_\rho \eta_{\mu\nu}\right)\right]
\label{F(k)}
\end{aligned}\ee
which is neither conserved nor traceless. 

Let us consider first $\tilde \EL$. We recall the Fourier transform
\be 
\int\! d^4x\, e^{ikx} \frac {1}{x^2} \log \mu^2 x^2 = \frac {4\pi^2 i}{k^2}
\left( \log 2-\gamma - \log \frac{k^2}{\mu^2}\right).
\label{Fxlogx}
\ee
Therefore, up to the term proportional to $(\log 2-\gamma)$, by Fourier
transforming (\ref{eq:2ptArbitraryDReg}) we obtain 
precisely (\ref{L(k)}) with $c= 1/\pi^4$, in agreement with the results of
\cite{Osborn93,Osborn96}. The term proportional 
to $(\log 2-\gamma)$ is to be added to (\ref{F(k)}). Now the divergence of
$\tilde \ET$ contains three independent terms 
proportional to $k^2 k_\nu k_\la k_\rho,\,\, k^4 k_\nu \eta_{\la\rho}$ and $k^4(
k_\la \eta_{\nu\rho}+ k_\rho \eta_{\nu\la})$, 
respectively, while the trace contains two independent terms proportional to
$k^2 k_\la k_\rho$ and $k^4 \eta_{\la\rho}$. 
On the other hand the ambiguity (\ref{eq:ambi}) contains the same 5 independent
terms with arbitrary coefficients. 
Therefore it is always possible to set to zero both the divergence and the trace
of $\tilde \ET$ by subtracting suitable counterterms. 
In the same way one can argue with the divergent term $\tilde \ED$. This term
deserves a comment: it is traceless and
divergenceless, but it is infinite, so it must be subtracted away along with the
$\tilde \EF$ term. Both $\EF$ and $\ED$, 
the Fourier anti-transforms of $\tilde{\EF}$ and $\tilde{\ED}$,  are contact
terms and they can be written in a compact form as
\be 
\langle\!\langle T_{\mu\nu}(x)\rangle\!\rangle &=& A'
\del_\mu\del_\nu\del_\lambda\del_\rho h^{\lambda\rho}(x)+
B'\left(\square \del_\mu \del_\lambda h^{\lambda}_\nu(x)+\square \del_\nu
\del_\lambda h^{\lambda}_\mu(x)\right) 
+ C' \eta_{\mu\nu} \square^2 h(x)\0\\&&+D' \,\square^2 h_{\mu\nu}(x)
+E'\left(\square \del_\mu\del_\nu h (x)
+\eta_{\mu\nu} \square\del_\lambda\del_\rho
h^{\lambda\rho}(x)\right)\label{Tmunu}
\ee
where $h= h_\lambda^\lambda$ and $A',B',C',D',E'$ are numerical coefficients
that contain also a part $\sim \frac 1{\delta}$. 
The local term to be subtracted from the action is proportional to
\be  
\int d^4x \left( \frac {A'}2 h^{\mu\nu} \del_ \mu\del_\nu\del_\lambda\del_\rho
h^{\lambda\rho}\right.
\!&+& \! B' h^{\mu\nu} \square \del_\mu \del_\lambda h^{\lambda}_\nu\0\\
&+&\left. \frac {C'}2 h \square^2 h +\frac {D'}2 h^{\mu\nu} \square^2
h_{\mu\nu}+E'  h^{\mu\nu} \square \del_\mu\del_\nu h\right) \label{counter}
\ee
We can conclude that the (regularized) Feynman diagram approach to the
$2$-point correlator is equivalent to regularizing the $2$-point function
calculated with the Wick
theorem approach. But we can draw also another, less pleasant, conclusion. Like
in $2d$, the Feynman diagrams
coupled to dimensional regularization may also produce unwelcome terms, such as
the $\ED$ and $\EF$ terms above, 
which must be subtracted away by hand. 

Finally we notice that, once (\ref{counter}) has been subtracted away, not only
the nonvanishing trace and divergence of the em tensor disappear, 
but the full contact term (\ref{Tmunu}) gets canceled. Thus the regularized
$2$-point correlator of the e.m. tensor coincides with the 
semiclassical expression.

\section{The $3$-point correlator}
\label{sec:3ptCorrelator}

The calculation of the $3$-point correlator brings new elements into the game.
First and foremost new (nontrivial) anomalies, 
but also an enormous complexity as compared to the $2$-point correlator.
In this section we first show that generically the $3$-point function of e.m.
tensors in $4d$ does not possess a parity-odd 
contribution due to the permutation symmetry of the correlator. Then we compute
the ``semiclassical'' $3$-point correlator by means 
of the Wick theorem in the same specific chiral 
fermionic model considered above, disregarding regularization. We find that, as
expected, the parity-odd part identically vanishes.
Subsequently we compute the same amplitude using Feynman diagrams and regularize
it. It turns out that not only the parity-even but also 
the parity-odd trace of the e.m. tensor is nonvanishing. We will explain this
apparent paradox in section \ref{sec:UglyDuckling}.

\subsection{No-go for parity-odd contributions}
\label{ssec:no-go}

In this subsection we will review the fact that in four dimensions there
are no parity-odd semiclassical contributions to the $3$-point function of
energy-momentum
tensors, which has already been emphasized in \citep{Zhiboedov,Stanev1}.

A very powerful tool to analyse which tensorial structures can exist
in a given correlation function in a CFT is the embedding formalism
as it was formulated in \citep{Costa:2011mg}. In their language,
to construct conformally covariant tensorial structures becomes a
game of putting together building blocks respecting the tensorial
requirements of your correlator. Particularly for the $3$-point function
of e.m. tensors we have seven building blocks. These building blocks
are written in terms of points $P_{i}$ of the six-dimensional embedding
space and lightlike polarization vectors $Z_{i}$. Three of them depend
on two points, namely
\begin{eqnarray}
H_{12} & = & -2\left[\left(Z_{1}\cdot Z_{2}\right)\left(P_{1}\cdot
P_{2}\right)-\left(Z_{1}\cdot P_{2}\right)\left(Z_{2}\cdot
P_{1}\right)\right],\label{eq:h12}\\
H_{23} & = & -2\left[\left(Z_{2}\cdot Z_{3}\right)\left(P_{2}\cdot
P_{3}\right)-\left(Z_{2}\cdot P_{3}\right)\left(Z_{3}\cdot
P_{2}\right)\right],\label{eq:h23}\\
H_{13} & = & -2\left[\left(Z_{1}\cdot Z_{3}\right)\left(P_{1}\cdot
P_{3}\right)-\left(Z_{1}\cdot P_{3}\right)\left(Z_{3}\cdot
P_{1}\right)\right].\label{eq:h13}
\end{eqnarray}
Four of them depend on three points, three being parity-even, namely
\begin{eqnarray}
V_{1} & = & \frac{\left(Z_{1}\cdot P_{2}\right)\left(P_{1}\cdot
P_{3}\right)-\left(Z_{1}\cdot P_{3}\right)\left(P_{1}\cdot
P_{2}\right)}{P_{2}\cdot P_{3}},\label{eq:v1}\\
V_{2} & = & \frac{\left(Z_{2}\cdot P_{3}\right)\left(P_{2}\cdot
P_{1}\right)-\left(Z_{2}\cdot P_{1}\right)\left(P_{2}\cdot
P_{3}\right)}{P_{3}\cdot P_{1}},\label{eq:v2}\\
V_{3} & = & \frac{\left(Z_{3}\cdot P_{1}\right)\left(P_{3}\cdot
P_{2}\right)-\left(Z_{3}\cdot P_{2}\right)\left(P_{3}\cdot
P_{1}\right)}{P_{1}\cdot P_{2}},\label{eq:v3}
\end{eqnarray}
while the last one is parity-odd, being the only object that one may
construct with an epsilon tensor, i.e.
\begin{equation}
O_{123}=\epsilon\left(Z_{1},Z_{2},Z_{3},P_{1},P_{2},P_{3}\right).\label{eq:o123}
\end{equation}

Our job now is to put together these objects to form a conformally
covariant object with the tensorial structure of the $3$-point function of
e.m. tensors. Particularly, the objects that we will construct must
present twice each polarization vector $Z_{i}$, since each $Z_{i}$
is associated with one index of the $i$-th e.m. tensor. Since we
are interested on parity-odd terms we will necessarily have the building
block $O_{123}$ which already takes care of one factor of each $Z_{i}$,
thus it is clear that our only options are
\begin{eqnarray}
T_{1} & = & O_{123}V_{1}V_{2}V_{3},\label{eq:T1}\\
T_{2} & = &
O_{123}\left(V_{1}H_{23}+V_{2}H_{13}+V_{3}H_{12}\right).\label{eq:T2}
\end{eqnarray}

In the following we will show that both $T_{1}$ and $T_{2}$ are
antisymmetric under the permutation of $1$ and $2$ for example,
which forbids them to be present in the $3$-point function of e.m. tensors.
By inspection of the expressions \eqref{eq:h12}-\eqref{eq:o123}
we see that under the exchange of $1$ and $2$ our building blocks
change as follows:
\begin{eqnarray*}
H_{12} & \rightarrow & H_{12},\\
H_{23} & \rightarrow & H_{13},\\
H_{13} & \rightarrow & H_{23},\\
V_{1} & \rightarrow & -V_{2},\\
V_{2} & \rightarrow & -V_{1},\\
V_{3} & \rightarrow & -V_{3},\\
O_{123} & \rightarrow & O_{123}.
\end{eqnarray*}
From these rules it is clear that both $T_{1}$ and $T_{2}$ are antisymmetric
under the exchange of $1$ and $2$. Of course the same result
holds for the exchanges $1\leftrightarrow3$ and $2\leftrightarrow3$. 

\subsection{The semiclassical parity-odd $3$-point correlator}
\label{ssec:ParityOddWick}

Consider a free chiral fermion $\psi_{L}$ in four dimensions which has the
$2$-point function\footnote{The factor of $\frac{1}{2\pi^2}$ in the propagator
of a fermion in $4d$ is needed in order for its Fourier-transform to give the
usual propagator, namely $\frac{i}{\slashed{p}}$.}
\begin{equation}
\left\langle
\psi_{L}\left(x\right)\overline{\psi_{L}}\left(y\right)\right\rangle 
=\frac{i}{2\pi^2}\frac{\gamma\cdot\left(x-y\right)}{\left(x-y\right)^{4}}P_{L},
\quad P_{L}=\frac{1-\gamma_{5}}{2},\label{eq:PropagatorChiralFermion}
\end{equation}
and the e.m. tensor
\begin{equation}
T_{\mu\nu}=\frac{i}{4}\left(\overline{\psi_{L}}\gamma_{\mu}\stackrel{
\leftrightarrow}{\partial}_{\nu}\psi_{L}
+\mu\leftrightarrow\nu\right),\text{ where
}\stackrel{\leftrightarrow}{\partial}_{\nu}\equiv\partial_{\nu}
-\stackrel{\leftarrow}{\partial}_{\nu}.\label{eq:EMTensorChiralFermion}
\end{equation}
Since we are dealing with a free theory we are able to compute the
$3$-point function of e.m. tensors by applying the Wick theorem. Using
the explicit form of the e.m. tensor \eqref{eq:EMTensorChiralFermion}
we write 
\begin{align}
\left\langle
T_{\mu\nu}\left(x\right)T_{\rho\sigma}\left(y\right)T_{\alpha\beta}
\left(z\right)\right\rangle 
 &\! =\! -\frac{i}{64}\left\langle \!\!\! 
\contraction{:}{\overline{\psi_{L}}}{\gamma_{\mu}\stackrel{\leftrightarrow}{
\partial}_{\nu}\psi_{L}:\left(x\right):
\overline{\psi_{L}}\gamma_{\rho}\stackrel{\leftrightarrow}{\partial}_{\sigma}
\psi_{L}: \left(y\right):
\overline{\psi_{L}}\gamma_{\alpha}\stackrel{\leftrightarrow}{\partial}_{\beta}}{
\psi_{L}} 
\bcontraction{:}{\overline{\psi_{L}}}{\gamma_{\mu}\stackrel{\leftrightarrow}{
\partial}_{\nu}\psi_{L}:\left(x\right):
\overline{\psi_{L}}\gamma_{\rho}\stackrel{\leftrightarrow}{\partial}_{\sigma}}{
\psi_{L}}
 :\overline{\psi_{L}}\gamma_{\mu}\stackrel{\leftrightarrow}{\partial}_{\nu} 
	\contraction{}{\psi_{L}}{:\left(x\right):}{\overline{\psi_{L}}}
	\bcontraction[2ex]{}{\psi_{L}}{:\left(x\right):\overline{\psi_{L}}
\gamma_{\rho}\stackrel{\leftrightarrow}{\partial}_{\sigma}\psi_{L}:
\left(y\right):}{\overline{\psi_{L}}}  
	\bcontraction[3ex]{\psi_{L}:\left(x\right):}{\overline{\psi_{L}}}{
\gamma_{\rho}\stackrel{\leftrightarrow}{\partial}_{\sigma}\psi_{L}:
\left(y\right):
\overline{\psi_{L}}\gamma_{\alpha}\stackrel{\leftrightarrow}{\partial}_{\beta}}{
\psi_{L}}
\psi_{L}:\left(x\right):
\overline{\psi_{L}}\gamma_{\rho}\stackrel{\leftrightarrow}{\partial}_{\sigma}
	\contraction{}{\psi_{L}}{:\left(y\right):}{\overline{\psi_{L}}}  
\psi_{L}: \left(y\right):
\overline{\psi_{L}}\gamma_{\alpha}\stackrel{\leftrightarrow}{\partial}_{\beta}
\psi_{L}:\left(z\right)\!\right\rangle \nonumber \\
 & \phantom{=}+\text{symmetrization}.\label{eq:ToWickContract}
\end{align}
There are two ways to fully contract these fields, as shown in equation
(\ref{eq:ToWickContract}). Each of the contractions is composed by a certain
tensor with six indices $f_{\nu a\sigma b\beta c}^{\left(i\right)}$ contracted
with a trace of six gamma matrices and a projector $P_{L}$, namely
\begin{equation}
f_{\nu a\sigma b\beta
c}^{\left(1\right)}\trace{\left(\gamma_{\mu}\gamma^{a}\gamma_{\rho}\gamma^{b}
\gamma_{\alpha}\gamma^{c}P_{L}\right)}
\text{ and }f_{\nu a\sigma b\beta c}^{\left(2\right)}
\trace{\left(\gamma_{\mu}\gamma^{a}\gamma_{\alpha}\gamma^{b}\gamma_{\rho}\gamma^
{c}P_{L}\right)},
\end{equation}
where the upper index of $f$ is $1$ for the first way of contracting and $2$ for
the second way.
The ordering of the free indices in the trace are given by the two
ways of performing the full contraction. The functions $f_{\nu a\sigma b\beta
c}^{\left(i\right)}$
are composed by eight terms which are the eight forms of distributing
the derivatives in the right hand side of \eqref{eq:ToWickContract}.
We will show that in reality $f^{\left(1\right)}$ and $f^{\left(2\right)}$
are the same object. To see this we will only need to exchange $a$
with $c$ in the expression for the second way of contracting, i.e.
\begin{equation}
f_{\nu a\sigma b\beta
c}^{\left(2\right)}\trace{\left(\gamma_{\mu}\gamma^{a}\gamma_{\alpha}\gamma^{b}
\gamma_{\rho}\gamma^{c}P_{L}\right)}
=f_{\nu a\sigma b\beta
c}^{\left(1\right)}\trace{\left(\gamma_{\mu}\gamma^{c}\gamma_{\alpha}\gamma^{b}
\gamma_{\rho}\gamma^{a}P_{L}\right)}.
\end{equation}
Hence, the sum of the two ways of contracting will simplify to
\begin{equation}
f_{\nu a\sigma b\beta
c}^{\left(1\right)}\left[\trace{\left(\gamma_{\mu}\gamma^{a}\gamma_{\rho}\gamma^
{b}\gamma_{\alpha}\gamma^{c}P_{L}\right)}
+\trace{\left(\gamma_{\mu}\gamma^{c}\gamma_{\alpha}\gamma^{b}\gamma_{\rho}
\gamma^{a}P_{L}\right)}\right].
\end{equation}
It is possible to put the second trace in the form 
$\trace{\left(\gamma_{\rho}\gamma^{a}\gamma_{\mu}\gamma^{c}\gamma_{\alpha}
\gamma^{b}P_{L}\right)}$, which reduces our final expression to
\begin{equation}
f_{\nu a\sigma b\beta
c}^{\left(1\right)}\left[\trace{\left(\gamma_{\mu}\gamma^{a}\gamma_{\rho}\gamma^
{b}\gamma_{\alpha}\gamma^{c}P_{L}\right)}
+\trace{\left(\gamma_{\rho}\gamma^{a}\gamma_{\mu}\gamma^{c}\gamma_{\alpha}
\gamma^{b}P_{L}\right)}\right].
\end{equation}
The trace of six gamma matrices and a gamma five is given by
\be\begin{aligned}
\trace{\left(\gamma_{\mu}\gamma_{a}\gamma_{\rho}\gamma_{b}\gamma_{\alpha}\gamma_
{c}\gamma_{5}\right)}= & \;
 4i\left(\eta_{\mu a}\epsilon_{\rho b\alpha c}-\eta_{\mu\rho}\epsilon_{ab\alpha
c}+\eta_{\rho a}\epsilon_{\mu b\alpha c}\right.\\
& \left. \;\; +\eta_{\alpha c}\epsilon_{\mu a\rho b}-\eta_{bc}\epsilon_{\mu
a\rho\alpha}+\eta_{\alpha b}\epsilon_{\mu a\rho c}\right).
\label{eq:Trace6gammas5}
\end{aligned}\ee
As one can easily check, the trace \eqref{eq:Trace6gammas5} is antisymmetric
under the exchange $\left(\mu\leftrightarrow\rho,b\leftrightarrow c\right)$,
thus the odd part of the correlation function is zero.

Now we will work out what are the functions $f^{\left(i\right)}$
and show the relation between $f^{\left(1\right)}$ and $f^{\left(2\right)}$
mentioned above. From the first way of contracting we derive the expression
\begin{equation}
\trace{\left[\gamma_{\mu}\partial_{\nu}\left(\gamma^{a}\partial_{a}\frac{1}{
\left(x-y\right)^{2}}P_{L}\right)\gamma_{\rho}
\partial_{\sigma}\left(\gamma^{b}\partial_{b}\frac{1}{\left(y-z\right)^{2}}P_{L}
\right)\gamma_{\alpha}\partial_{\beta}
\left(\gamma^{c}\partial_{c}\frac{1}{\left(z-x\right)^{2}}P_{L}\right)\right]}
+\cdots,\label{eq:FirstFullContraction}
\end{equation}
where the ellipsis stand for the seven other ways of organizing the
derivatives $\partial_{\nu}$, $\partial_{\sigma}$ and $\partial_{\beta}$.
From \eqref{eq:FirstFullContraction} we see that we will have some
expression that we call $f^{\left(1\right)}$ contracted with
$\trace{\left(\gamma_{\mu}\gamma^{a}\gamma_{\rho}\gamma^{b}\gamma_{\alpha}
\gamma^{c}P_{L}\right)}$.
The expression for $f^{\left(1\right)}$ can be read off from
\eqref{eq:FirstFullContraction}:
\small
\begin{eqnarray}
f_{\nu a\sigma b\beta c}^{\left(1\right)} & = &
\partial_{\nu}^{x}\partial_{a}^{x}\frac{1}{\left(x-y\right)^{2}}\partial_{\sigma
}^{y}\partial_{b}^{y}\frac{1}{\left(y-z\right)^{2}}\partial_{\beta}^{z}\partial_
{c}^{z}\frac{1}{\left(z-x\right)^{2}}-\partial_{\sigma}^{y}\partial_{a}^{x}\frac
{1}{\left(x-y\right)^{2}}\partial_{\beta}^{z}\partial_{b}^{y}\frac{1}{
\left(y-z\right)^{2}}\partial_{\nu}^{x}\partial_{c}^{z}\frac{1}{
\left(z-x\right)^{2}}\nonumber \\
 &  &
-\partial_{\sigma}^{y}\partial_{\nu}^{x}\partial_{a}^{x}\frac{1}{
\left(x-y\right)^{2}}
\left[\partial_{b}^{y}\frac{1}{\left(y-z\right)^{2}}\partial_{\beta}^{z}
\partial_{c}^{z}\frac{1}{\left(z-x\right)^{2}}-
\partial_{\beta}^{z}\partial_{b}^{y}\frac{1}{\left(y-z\right)^{2}}\partial_{c}^{
z}\frac{1}{\left(z-x\right)^{2}}\right]\nonumber \\
 &  &
-\partial_{\beta}^{z}\partial_{\sigma}^{y}\partial_{b}^{y}\frac{1}{
\left(y-z\right)^{2}}
\left[\partial_{\nu}^{x}\partial_{a}^{x}\frac{1}{\left(x-y\right)^{2}}\partial_{
c}^{z}\frac{1}{\left(z-x\right)^{2}}
-\partial_{a}^{x}\frac{1}{\left(x-y\right)^{2}}\partial_{\nu}^{x}\partial_{c}^{z
}\frac{1}{\left(z-x\right)^{2}}\right]\nonumber \\
 &  &
-\partial_{\nu}^{x}\partial_{\beta}^{z}\partial_{c}^{z}\frac{1}{
\left(z-x\right)^{2}}
\left[\partial_{a}^{x}\frac{1}{\left(x-y\right)^{2}}\partial_{\sigma}^{y}
\partial_{b}^{y}\frac{1}{\left(y-z\right)^{2}}
-\partial_{\sigma}^{y}\partial_{a}^{x}\frac{1}{\left(x-y\right)^{2}}\partial_{b}
^{y}\frac{1}{\left(y-z\right)^{2}}\right].
\end{eqnarray}
\normalsize
The second way of contracting give us the expression
\begin{equation}
\trace{\left[\gamma_{\mu}\partial_{\nu}\left(\gamma^{a}\partial_{a}\frac{1}{
\left(x-z\right)^{2}}P_{L}\right)
\gamma_{\alpha}\partial_{\beta}\left(\gamma^{b}\partial_{b}\frac{1}{
\left(z-y\right)^{2}}P_{L}\right)\gamma_{\alpha}\partial_{\beta}
\left(\gamma^{c}\partial_{c}\frac{1}{\left(y-x\right)^{2}}P_{L}\right)\right]}
+\cdots,\label{eq:SecondtFullContraction}
\end{equation}
from where we may read off the expression for $f^{\left(2\right)}$:
\small
\begin{eqnarray}
f_{\nu a\sigma b\beta c}^{\left(2\right)} & = &
\partial_{\nu}^{x}\partial_{a}^{x}
\frac{1}{\left(x-z\right)^{2}}\partial_{\beta}^{z}\partial_{b}^{z}\frac{1}{
\left(z-y\right)^{2}}\partial_{\sigma}^{y}\partial_{c}^{y}
\frac{1}{\left(y-x\right)^{2}}-\partial_{\beta}^{z}\partial_{a}^{x}\frac{1}{
\left(x-z\right)^{2}}\partial_{\sigma}^{y}\partial_{b}^{z}
\frac{1}{\left(z-y\right)^{2}}\partial_{\nu}^{x}\partial_{c}^{y}\frac{1}{
\left(y-x\right)^{2}}\nonumber \\
 &  &
-\partial_{\nu}^{x}\partial_{\sigma}^{y}\partial_{c}^{y}\frac{1}{
\left(y-x\right)^{2}}\left[\partial_{a}^{x}\frac{1}{\left(x-z\right)^{2}}
\partial_{\beta}^{z}\partial_{b}^{z}\frac{1}{\left(z-y\right)^{2}}-\partial_{
\beta}^{z}\partial_{a}^{x}\frac{1}{\left(x-z\right)^{2}}
\partial_{b}^{z}\frac{1}{\left(z-y\right)^{2}}\right]\nonumber \\
 &  &
-\partial_{\sigma}^{y}\partial_{\beta}^{z}\partial_{b}^{z}\frac{1}{
\left(z-y\right)^{2}}\left[\partial_{\nu}^{x}\partial_{a}^{x}
\frac{1}{\left(x-z\right)^{2}}\partial_{c}^{y}\frac{1}{\left(y-x\right)^{2}}
-\partial_{a}^{x}\frac{1}{\left(x-z\right)^{2}}\partial_{\nu}^{x}
\partial_{c}^{y}\frac{1}{\left(y-x\right)^{2}}\right]\nonumber \\
 &  &
-\partial_{\beta}^{z}\partial_{\nu}^{x}\partial_{a}^{x}\frac{1}{
\left(x-z\right)^{2}}\left[\partial_{b}^{z}\frac{1}{\left(z-y\right)^{2}}
\partial_{\sigma}^{y}\partial_{c}^{y}\frac{1}{\left(y-x\right)^{2}}-\partial_{
\sigma}^{y}\partial_{b}^{z}\frac{1}{\left(z-y\right)^{2}}
\partial_{c}^{y}\frac{1}{\left(y-x\right)^{2}}\right].
\end{eqnarray}
\normalsize
It is now a straightforward exercise to check that if one exchanges
$a$ with $c$ in the expression of $f_{\nu a\sigma b\beta c}^{\left(2\right)}$
one gets $f_{\nu a\sigma b\beta c}^{\left(1\right)}$, i.e.
\begin{equation}
f_{\nu c\sigma b\beta a}^{\left(2\right)}=f_{\nu a\sigma b\beta
c}^{\left(1\right)}.
\end{equation}

\subsection{Relevant Fourier transforms}
\label{ssec:Fourier}

In the next subsection, in order to compute the $3$-point amplitude of the e.m.
tensor, with the Feynman diagram technique we will use (momentum space) Feynman
diagrams. Although essentially equivalent to the Wick theorem they lend
themselves more naturally to regularization. The two techniques are related by
Fourier transform. Hereby we collect a series of Fourier transforms of
distributions that are used in our calculations. The source is \cite{Gelfand}.
The notation is as follows
\be 
 {\cal F}[\phi(x)](k)\equiv \tilde\phi(k) = \int d^4x \, e^{ikx}
\phi(x),\quad\quad \phi(x) = \int\frac {d^4k}{(2\pi)^4} 
e^{-ikx}\tilde \phi(k)\0
\ee
In particular 
\be
\int d^4 x \, e^{ikx}\,\frac{1}{x^2} &=& \frac{4\pi^2 i}{k^2},\\
\int d^4 x \, e^{ikx}\,\frac{\log x^2\mu^2}{x^2} &=& -\frac{4\pi^2
i}{k^2}\log\left(\frac{-k^2}{\bar{\mu}^2}\right),
\ee
where $\bar{\mu}^2\equiv 2\mu^2 e^{-\gamma}$, $\gamma=0.57721\dots$ being the
Euler constant. As we have seen this is essentially what one needs to compute
the Fourier transform of the $2$-point correlator. The novel feature in the
calculation of the $3$-point correlator is the appearance of products of similar
expressions in different points, a prototype being
\be 
\frac{1}{(x-y)^2 (x-z)^2 (y-z)^2}.\label{basic}
\ee
This is singular at coincident points and has a non-integrable singularity at
$x=y=z=0$.
Ignoring this let us proceed to Fourier-transforming it 
\begin{align}
&\int d^4x\, d^4y\, d^4z\,  \frac{e^{i(k_1 x+k_2 y-qz)} }{(x-y)^2 (x-z)^2
(y-z)^2}=
\int d^4x \, d^4y\, d^4z\, \frac{e^{i(k_1 x+k_2 y+(k_1+k_2-q)z)}}{(x-y)^2 x^2
y^2}\0\\
&=(2\pi)^4 \delta(q-k_1-k_2)\int d^4x \, d^4y\, \frac{e^{i(k_1 x+k_2y)}
}{(x-y)^2 x^2 y^2}.\label{B1}
\end{align}
Let us set $f(x,y) = \frac 1 {(x-y)^2 x^2 y^2}$. Then, using the convolution
theorem, the Fourier transform of $f$ with respect to $x$ is
\be
{\cal F}_x[f(x,y)](k_1)&=& \int d^4x\, e^{ik_1x}f(x,y) =\frac 1{y^2} \int d^4x\,
\frac  {e^{ik_1x}}{x^2(x-y)^2}\0\\
&=& \frac 1{y^2} \int \frac{d^4p}{(2\pi)^4} \, {\cal F}_x\left[\frac
1{x^2}\right]\!(k_1-p)\,  
{\cal F}_x\left[\frac 1{(x-y)^2}\right]\!(p)\0\\
&=& -\frac 1{y^2} \int \! d^4p \,\frac {e^{ipy}}{p^2(p-k_1)^2}.\label{B2}
\ee
Therefore
\be
\int\! d^4x \, d^4y\, \frac{e^{i(k_1 x+k_2y)} }{(x-y)^2 x^2 y^2}
&=&\! \int d^4y \,e^{ik_2y}\, {\cal F}_x[f(x,y)](k_1)\0\\
&=&-i(2\pi)^6\int\!\frac{d^4p}{(2\pi)^4}\, \frac
1{p^2(p-k_1)^2(p+k_2)^2}.\label{B3}
\ee
We can now compute the RHS of (\ref{B3}) in the usual way by introducing a Feynman
parametrization in terms of two parameters $u,v$:
\be 
&&\int\!\frac{d^4p}{(2\pi)^4}\, \frac 1{p^2(p-k_1)^2(p+k_2)^2} =\int_0^1\! du\int_0^{1-u} \! dv 
\int\!\frac{d^4p'}{(2\pi)^4} \,\frac
1{({p'}^2-\ell^2+\Delta)^3}
\label{B4}
\ee
where $p'=p-u k_1+v k_2$ and $\Delta = u(1-u) k_1^2+v(1-v) k_2^2 + 2 uv
\,k_1k_2$.  Performing the $p'$ integral one gets 
\be 
\int_0^1\! du\int_0^{1-u} \! dv\int\!\frac{d^4p}{(2\pi)^4}\int\!\frac{d^\delta
\ell}{(2\pi)^\delta}\frac 1{({p}^2-\ell^2+\Delta)^3} = \frac i{2(4\pi)^2 } 
\int_0^1\! du\int_0^{1-u} \! dv\frac 1{\Delta}\label{basicreg}
\ee
Our attitude will be to define the regularization of (\ref{basic}) as the Fourier 
anti-transform of the (\ref{basicreg}). 
 
In general, however, the expressions we have to do with are not as simple as 
(\ref{basicreg}) and the integrals as simple as (\ref{B3}). The typical integral 
of the type (\ref{B3}) contains a polynomial of $p, k_1,k_2$ in the numerator
of the integrand. In this case we have two ways to proceed: either we extend the
running momentum $p$ to extra dimensions (dimensional regularization), as we have done in 2d,
carry out the integration and Fourier-anti-transform the final result, 
or we reduce the calculations to a differential operator applied to the Fourier-anti-transform
of (\ref{basicreg}) (differential regularization). Usually the former procedure is more convenient, 
while in many cases the latter is problematic.

Other analogous expressions are obtained in appendix \ref{sec:Fourier}.

\subsection{The parity-odd $3$-point correlator with Feynman diagrams}
\label{ssec:3ptsodd}

This section is devoted to the same calculation as in subsection
(\ref{ssec:ParityOddWick}), but with Feynman diagram techniques. In order to
compute the $3$-point function of the energy-momentum tensor for a chiral
fermion, it is very convenient to couple it minimally to gravity and extract
from the corresponding action the Feynman rules, as in \cite{DS,BGL}. The relevant
formalism and notation is reviewed in appendix \ref{sec:chiralm}. Due to the
non polynomial character of the action the diagrams contributing to the trace
anomaly are infinitely many. Fortunately, using diffeomorphism invariance, it is
enough to determine the lowest order contributions and consistency takes care of
the rest.
There are two potential lowest order diagrams that may contribute. The first
contribution, the bubble graph, turns out to give a vanishing contribution. The
important term comes from the triangle graph. This has an incoming line with
momentum $q=k_1+k_2$ with Lorentz indices $\mu,\nu$. The two outgoing lines have
momenta $k_1,k_2$ with Lorentz labels $\lambda,\rho$ and $\alpha,\beta$,
respectively. The contribution is formally written as
\begin{align} 
&\ET^{(1)}_{\mu\nu\alpha\beta\lambda\rho}(k_1,k_2) =-\frac 1 {512}\int \frac
{d^4p}{(2\pi)^4}\, {\rm tr} \left[\left(\frac
1{\slashed{p}}\bigl((2p-k_1)_\lambda
\gamma_\rho+(\lambda\leftrightarrow \rho)\bigr)\right.\right.\frac
1{\slashed{p}-\slashed{k}_1}\label{T1}\\
& \times \bigl((2p-2 k_1 -
k_2)_{\al}\gamma_{\beta}+(\al\leftrightarrow \beta)\bigr)
\left.\left.\frac 1{\slashed{p} - \slashed{q}}\bigl( (2{p}
-{q})_\mu\gamma_\nu+(\mu\leftrightarrow \nu)\bigr)\right) \frac
{1+\gamma_5}2\right] \0
\end{align}
to which the cross graph contribution
$\ET^{(2)}_{\mu\nu\alpha\beta\lambda\rho}(k_1,k_2)=\ET^{(1)}_{
\mu\nu\lambda\rho\alpha\beta}(k_2,k_1)$ has to be added. We regularize
(\ref{T1}) as usual by introducing extra component of the
momentum running around the loop $p\to p+\ell$,
$\ell=\ell_4,\ldots,\ell_{\delta+4}$:
\be
&&\ET^{(1)}_{\mu\nu\alpha\beta\lambda\rho }(k_1,k_2)
= -\frac 1 {512} \int \frac {d^4p}{(2\pi)^4}
\int \frac {d^{\delta}\ell}{(2\pi)^{\delta}} \,\tr\left( \frac
{\slashed{p}+\slashed{\ell}}{p^2-\ell^2} (2p-k_1)_\lambda\gamma_\rho
\,\right.\0\\
&&\times \,\frac {\slashed{p}+\slashed{\ell}-\slashed{k}_1}{(p-k_1)^2-\ell^2}
\,(2p-2k_1-k_2)_{\al}\gamma_{\beta} \left.\frac
{\slashed{p}+\slashed{\ell}-\slashed{q}}{(p-q)^2-\ell^2}\, (2{p}
-{q})_\mu\gamma_\nu
\frac {1+\gamma_5}2 \right)\label{T2}
\ee
where the symmetrization with respect to $\al\leftrightarrow \beta$, $\lambda
\leftrightarrow \rho$ and $\mu\leftrightarrow \nu$ is understood. 
We should now proceed to the explicit calculation. However one quickly realizes
that
this involves a huge number of terms. To find an orientation among the latter it
is very
useful to first compute the trace and the divergence of the e.m. tensor in the
above formulas.
They are connected to the trace and divergence of the full one-loop e.m. tensor
by the 
general formulas of section \ref{ssec:genform}.

\subsubsection{The trace}

The trace of (\ref{T2}) is
\be
&&{\ET^{(1a)}}^\mu_{\mu\alpha\beta\lambda\rho }(k_1,k_2)
= -\frac 1 {256} \int \frac {d^4p}{(2\pi)^4}
\int \frac {d^{\delta}\ell}{(2\pi)^{\delta}} \,\tr\left( \frac
{\slashed{p}+\slashed{\ell}}{p^2-\ell^2} (2p-k_1)_\lambda\gamma_\rho
\,\right.\0\\
&&\times\left. \,\frac
{\slashed{p}+\slashed{\ell}-\slashed{k}_1}{(p-k_1)^2-\ell^2}
\,(2p-2k_1-k_2)_{\alpha}\gamma_{\beta} \frac
{\slashed{p}+\slashed{\ell}-\slashed{q}}{(p-q)^2-\ell^2}\, (2\slashed{p}
-\slashed{q}) 
\frac {1+\gamma_5}2 \right).\label{T3}
\ee
On the other hand if we first take the trace of (\ref{T1}) and then regularize
it, we get
\be
&&{\ET^{(1b)}}^\mu_{\mu\alpha\beta\lambda\rho }(k_1,k_2)
= -\frac 1 {256} \int \frac {d^4p}{(2\pi)^4}
\int \frac {d^{\delta}\ell}{(2\pi)^{\delta}} \,\tr\left( \frac
{\slashed{p}+\slashed{\ell}}{p^2-\ell^2} (2p-k_1)_\lambda\gamma_\rho
\,\right.\0\\
&&\times \left. \frac
{\slashed{p}+\slashed{\ell}-\slashed{k}_1}{(p-k_1)^2-\ell^2}
\,(2p-2k_1-k_2)_{\al}\gamma_{\beta}\frac{\slashed{p}
+\slashed{\ell}-\slashed{q}}{(p-q)^2-\ell^2}\, \left(2\slashed{p}+2
\slashed{\ell} -\slashed{q}\right) 
\frac {1+\gamma_5}2 \right).\label{T4}
\ee
The difference between the two is\footnote{Eqs.(\ref{reconstruction}) and
(\ref{2ndtrace}) suggest that the right prescription is (\ref{T4}), not
(\ref{T3}). This has been fully confirmed by the calculations in $2d$. The
anomaly is determined by the $n$-point functions 
where the entries are one trace of the e.m. tensor
and $n-1$ e.m. tensors. We have quoted the `wrong' formula (\ref{T3}) on purpose
in order to stress this point.}
\be
&&\Delta {\ET^{(1)}}^\mu_{\mu\alpha\beta\lambda\rho }(k_1,k_2)
= -\frac 1 {128} \int \frac {d^4p}{(2\pi)^4}
\int \frac {d^{\delta}\ell}{(2\pi)^{\delta}} \,\tr\left( \frac
{\slashed{p}+\slashed{\ell}}{p^2-\ell^2} (2p-k_1)_\lambda\gamma_\rho
\,\right.\0\\
&&\times \left.\,\frac
{\slashed{p}+\slashed{\ell}-\slashed{k}_1}{(p-k_1)^2-\ell^2}
\,(2p-2k_1-k_2)_{\al}\gamma_{\beta}  \frac
{\slashed{p}+\slashed{\ell}-\slashed{q}}{(p-q)^2-\ell^2}\,   \slashed{\ell} \, 
\frac {1+\gamma_5}2 \right).\label{DeltaT}
\ee
Similar expressions hold for $\ET^{(2)}$. Now it is easy to show that (\ref{T3})
vanishes along with the analogous expression for $\ET^{(2)}$, while  (\ref{T4})
does not, and in fact the odd-parity part of (\ref{DeltaT}) is precisely the
anomalous term computed in \cite{BGL}, which, together
with the cross term coming form $\ET^{(2)}$, gives rise to the Pontryagin
anomaly. More precisely, the two terms yield
\be 
\ET^{\mu}_{\mu\alpha\beta\lambda\rho }(k_1,k_2)=\frac 1{192(4\pi)^2} k_1^\sigma
k_2^\tau\left(t^{(21)}_{\lambda\rho\al\beta\sigma\tau}-
t_{\lambda\rho\al\beta\sigma\tau} (k_1^2+k_2^2+k_1k_2)\right)\label{Tmumu}
\ee
The tensors $t$ and $t^{(21)}$ were defined in \cite{BGL}. 
In \cite{BGL} the external lines were put on shell (in the de Donder gauge):
$k_1^2=k_2^2=0$. This is the right thing to do, as we shall see, but it is
important to clarify the role of the off-shell terms too. Therefore let us
consider nonvanishing external square momenta. While the remaining terms,
when inserted into the reconstruction formula (\ref{reconstruction}), reproduce
the Pontryagin density to order $h^2$,
\be
\sim\epsilon^{\mu\nu\lambda \rho} \left(\del_\mu\del_\sigma h^\tau_\nu \,
\del_\lambda\del_\tau h_{\rho}^\sigma-
\del_\mu\del_\sigma h^\tau_\nu \, \del_\lambda\del^\sigma h_{\tau\rho}\right)+
{\cal O}(h^3),\label{Pont2ndorder}
\ee
the term proportional to $k_1^2+k_2^2$ in (\ref{Tmumu}) leads to a term
proportional to
\be
\epsilon^{\mu\nu\lambda \rho} \del_\mu \square h_\nu^{\alpha}\del_\lambda
h_{\rho\alpha}.\label{noncovariant}
\ee
They are both invariant under the Weyl rescaling $\delta h_{\mu\nu}= 2\omega
\,\eta_{\mu\nu}$. Thus the corresponding anomalous terms
obtained by integrating (\ref{Pont2ndorder}) and (\ref{noncovariant}) multiplied
by the Weyl parameter $\omega$ are consistent. But while
the first gives rise to a true anomaly, the second one must be trivial because
there is no covariant cocycle containing the
$\epsilon$ tensor beside the Pontryagin one. In fact it is easy to guess the
counterterm that cancels it: it is proportional to
\be
\int d^4x\, h\, \epsilon^{\mu\nu\lambda \rho} \del_\mu \square
h_\nu^{\alpha}\del_\lambda h_{\rho\alpha}\label{counterterm}
\ee
where $h=h_\mu^\mu$.
But this counterterm breaks invariance under general coordinate transformations,
which to lowest order take the form 
$\delta_\xi h_{\mu\nu}=\del_\mu \xi_\nu+\del_\nu\xi_\mu$ (with $\delta_\xi
\omega=0$). Thus we must expect that off-shell terms
break the e.m. tensor conservation. This does not mean that there are true diff
anomalies, but simply that we have to subtract 
counterterms (actually, a lot of them, see below) in order to recover a
covariant regularization. 
In other words taking into account off-shell terms is a very effective way to
complicate one's own life, while disregarding them 
does not spoil the result if our aim is to find a covariant expression of the
anomaly. The reason for this is that  the
equation of motion of gravity in vacuum
\be 
\square h_{\mu\nu}-\d_{\mu} \d_{\lambda}h^\lambda _\nu - 
\d_\nu \d_{\lambda} h^\lambda _\mu +\d_\mu\d_\nu
h^\lambda_\lambda=0\label{linrmunu}
\ee
is covariant. If we impose the De Donder gauge
\be
2\d_\mu h^{\mu}_\lambda -\d_\lambda h^\mu_\mu=0\label{lindedonder}
\ee
the last three terms in the RHS of (\ref{linrmunu}) vanish and the latter
reduces to $\square h_{\mu\nu}= 0$. Therefore choosing this gauge and
putting the external legs on shell (as we have just done) does not break
covariance and considerably simplifies the calculations\footnote{Sometimes it
oversimplifies them, for instance in $2d$ or in $4d$ for the $2$-point
correlator. In such cases there is no way but doing the calculations in full, as
we have done above.}.

\subsubsection{The divergence}
\label{ssec:divergence}

The discussion in the previous subsection raises a problem. For not only can we
subtract (\ref{noncovariant}) via the counterterm (\ref{counterterm}), but also
(\ref{Pont2ndorder}) can be subtracted away by means of the counterterm
\be
 \sim \int d^4x\, h\,\epsilon^{\mu\nu\lambda \rho} \left(\del_\mu\del_\sigma
h^\tau_\nu \, \del_\lambda\del_\tau h_{\rho}^\sigma-
\del_\mu\del_\sigma h^\tau_\nu \, \del_\lambda\del^\sigma h_{\tau\rho}\right)+
{\cal O}(h^3),\label{Pont2ndordercount}
\ee
as it is easy to verify. This of course generates new terms in the divergence of
the e.m. tensor. Choosing the on-shell option to simplify the problem, they
corresponds, in the momentum notation, to the terms
\be 
\sim \epsilon_{\beta\rho\sigma\tau}\, k_{1\nu} k_1^\sigma  k_2^\tau
\left(k_{1\lambda} k_{2\alpha}- \eta_{\al\lambda} k_1\cdot k_2\right)+
\{\lambda\leftrightarrow \rho\}+ \{\alpha  \leftrightarrow \beta\} +\{ 1
\leftrightarrow 2\}\label{deltadiff}
\ee
where the subscript $\nu$, in coordinate representation, is saturated with the
diffeomorphism parameter $\xi^\nu$. 

Let us remark that, when we refer to the lowest order in $h$, any anomaly
appears to be trivial and can be subtracted  (see what we have done above in
$2d$). This is true also for the even parity anomalies, but it is an accident of
the approximation. 
What is decisive about triviality or not of the anomalies is their diff partner.
We must arrive at a configuration in which the diff partner of the trace anomaly
vanishes. In this case we can conclude that a nonvanishing trace anomaly is
nontrivial even if it is expressed at the lowest order in $h$. This expression
will be the lowest order expansion of a covariant expression (much as
(\ref{Pont2ndorder}) is). In conclusion we expect that subtracting away
(\ref{Pont2ndorder}) by means of (\ref{Pont2ndordercount}) is a forbidden
operation (it breaks covariance). But it is important to verify it by a direct
calculation. This is what we intend to do in the sequel.

The relevant lowest order contribution to $\langle\!\langle \nabla^\mu
T_{\mu\nu}\rangle\!\rangle$, see (\ref{2nddivergence}), comes from the $3$-point
function $\langle 0|\mathcal{T}\{\del^\mu T_{\mu\nu}
(x)T_{\lambda\rho}(y)T_{\alpha\beta}(z)\}|0\rangle$. The latter 
corresponds to two graphs, the bubble and the triangle ones (see \cite{BGL}).
The bubble graph contribution vanishes. 
The triangle contribution is given by
\be 
&& q^\mu \ET^{(1)}_{\mu\nu\lambda\rho\alpha\beta}(k_1,k_2) =-\frac 1 {512}\int
\frac
{d^4p}{(2\pi)^4}\, {\rm tr} \left[\left(\frac
1{\slashed{p}}\bigl((2p-k_1)_\lambda
\gamma_\rho+(\lambda\leftrightarrow \rho)\bigr)\right.\right.\frac
1{\slashed{p}-\slashed{k}_1}\label{Div1}\\
&& \times \bigl((2p-2 k_1 -
k_2)_{\al}\gamma_{\beta}+(\al\leftrightarrow \beta)\bigr)
\left.\left.\frac 1{\slashed{p} - \slashed{q}}\bigl( (2{p}
-{q})\cdot q\,\gamma_\nu+ (2{p}-{q})_\nu \slashed{q}\bigr)\right) \frac
{1+\gamma_5}2\right] \0
\ee
to which the cross contribution $q^\mu
\ET^{(1)}_{\mu\nu\alpha\beta\lambda\rho}(k_2,k_1)$ has to be added. We regulate
the integral as usual
with an extra dimensional momentum $\ell$ and introduce Feynman parameters as
needed. After a rather lengthy algebra, in particular
with explicit use of the identity
\be
\eta_{\mu\nu} \epsilon_{\la\rho\sigma\tau} 
-\eta_{\mu\la}\epsilon_{\nu\rho\sigma\tau}+\eta_{\mu\rho}\epsilon_{
\nu\la\sigma\tau}-
\eta_{\mu\si}\epsilon_{\nu\la\rho \tau} +\eta_{\mu\tau} \epsilon_{\nu\la\rho
\si}=0,\label{idgamma}
\ee
the regularized (\ref{Div1}) can be recast into the form
\be
&&\ED^{(1)}_{\nu\la\rho\al\beta} (k_1,k_2)\equiv q^\mu
\left(\ET^{(1)}_{\mu\nu\lambda\rho\alpha\beta}(k_1,k_2)+\ET^{(1)}_{
\mu\nu\alpha\beta\lambda\rho}(k_2,k_1)\right) \0\\
&&=\frac i {256} \int_0^1\!dx \int_0^{1-x}\!dy\, \int \frac
{d^4p}{(2\pi)^4}\int \frac
{d^\delta \ell}{(2\pi)^\delta} \biggl[- \epsilon_{\nu\beta\si\tau}\left(p_\rho
k_1^\si k_2^\tau +(k_{1\rho} k_2^\tau + k_{2\rho}k_1^\tau+2k_{2}^\tau p_\rho)
p^\si\right)\0\\
&&\quad + \epsilon_{\nu\beta\rho\tau}\left(p^2(k_1+k_2 -p)^\tau+ p\!\cdot\! k_1
\,k_2^\tau - p\!\cdot\! k_2 \,k_1^\tau+  
(k_2  -2p)\!\cdot \!k_1 p^\tau\right)+\epsilon_{\nu\si\tau\kappa}
\eta_{\beta\rho}\,p^\si k_1^\tau k_2^\kappa\0\\
&&\quad +\epsilon_{\nu\rho\si\tau} \left(p_\beta k_1^\tau k_2^\si
+p^\si(k_{1\beta} k_2^\tau + k_{2\beta}k_1^\tau-2p_\beta k_1^\tau) \right) 
+     \ell^2   \epsilon_{\nu\beta\rho\tau} (p+k_1-k_2)^\tau\biggr]\0\\
&&\quad \times \,\frac {2 p\cdot\! (k_1+k_2) (2p+k_1)_\la (2p-k_2)_\al}{\left[
(p+xk_1-yk_2)^2+2xy
 \, k_1\!\cdot\! k_2-\ell^2\right]^3}.\label{Div2}
\ee
This expression does not contain any of the terms (\ref{deltadiff}), but of
course this is not enough. We have to prove that all the terms in (\ref{Div2})
either vanish or are trivial in the sense that they can be canceled by
counterterms that are Weyl invariant.
This analysis is carried out in appendix \ref{sec:divergence}, where
counterterms are constructed which cancel all the 
nonvanishing terms in (\ref{Div2}) without altering the result of the trace
anomaly calculation. Thus the lowest order expression
(\ref{Pont2ndorder}) cannot be canceled (except at the price of breaking
diffeomorphism invariance) and is a genuine covariant expression.
It represents the lowest order approximation of the Pontryagin density.

\subsubsection{(Partial) conclusion}

The results obtained in this section fully confirm those of
\cite{ChD1,ChD2,BGL}. The apparent contradiction inherent in the fact that the 
semiclassical parity-odd correlator of three energy-momentum tensors vanishes
will be explained in the next section. Here we would like to 
draw some conclusion on the regularized e.m. tensor $3$-point function. We have
seen that the trace and the traceless part of the correlator 
must be regularized separately.
The traceless part of the correlator can be regularized starting from
(\ref{T2}). We would like to be able to conclude that the regularized 
traceless part coincides with the semiclassical part, i.e. it vanishes, but in
order to justify this conclusion the calculations are very 
challenging, because it is not enough to regularize and compute (\ref{T2}), but
we must also take into account all the counterterms (with 
the exact coefficients) that we have subtracted in order to guarantee
covariance, see appendix \ref{sec:divergence}. This can realistically 
be done only with a computer algebra program. For the time being, although we
believe the regularized traceless part of the correlator 
vanishes, we leave its proof as an open problem. 

Finally a comment on the parity-even part of the $3$-point e.m. tensor
correlator. The calculation of the trace and divergence involves 
many more terms than in the parity-odd part, but it does not differ in any
essential way from it. Also in the parity-even part it is 
necessary to introduce  counterterms in order to guarantee covariance and the
correct final expression for the trace anomaly. On the 
other hand this is pretty clear already in the $2d$ case, as we have shown
above. Since the results for the parity-even part of the 
$3$-point function, both semiclassical and regularized, \cite{Osborn93,Osborn96}, 
and relevant even-parity anomalies are well-known, see \cite{Duff}, 
we dispense with an explicit calculation.

\section{The ugly duckling anomaly}
\label{sec:UglyDuckling}
 
The title is due to the non-overwhelming consideration met so far by the
Pontryagin trace anomaly.
Needless to say its presence in the free chiral fermion model is at first sight
surprising. The basic ingredient to evaluate this 
anomaly in the Feynman diagram approach is traditionally the triangle diagram,
which can be seen as the lowest order approximation of 
the $3$-point correlator, whose entries are one e.m. trace and two e.m.
tensors. 
On the other hand, since the semiclassical parity-odd part of the $3$-point
correlator of the e.m. tensor vanishes on the basis of 
very general considerations of symmetry, it would seem that even the triangle
diagram contributions should vanish, because the 
regularization of zero should be zero.

The remark made in connection with formulas (\ref{T3}), (\ref{T4}) and
(\ref{DeltaT}) may seem to add strength
to this argument because it leaves the impression that the Pontryagin anomaly is
something we can do without. After all its existence in the $3$-point
correlators is related to the order in which we regularize. One might argue that
if we regularize in a specific order the anomaly disappears, but this is not the
case. First of all we remark that what one does in all kind of anomalies is to
regularize the divergence of a current or of the e.m. tensor, or the trace of
the latter, rather than regularizing the current or the e.m. tensor and then
taking the divergence or the trace thereof. In other words the regularization
should be done independently for each irreducible component that enters into
play. But, even forgetting this, in order to make a decision about such an
ambiguous occurrence one must resort to some consistency argument, and this is
what we will do below.

In fact the apparent contradiction is based on a misunderstanding, which
consists in assuming that the (unregulated) $3$-point correlator in the
coordinate representation is the sole ingredient of the anomaly. This is not
true\footnote{We remark that the parity-even  $3$-point correlator of the e.m.
trace and two e.m. tensors also vanishes semiclassically, but this does not
prevent the even parity anomaly from being nonvanishing.}. The  
$3$-point correlator of the energy-momentum tensor is one of the possible
markers of the trace anomaly, but, as we shall see, there are infinite many of
them and consistency demands that they all agree (the more so if the correlator
is unregulated). Let us start with by clarifying this point.

In subsection \ref{ssec:genform} we have shown how to reconstruct the full
one-loop e.m. tensor starting from the one-loop correlators of the e.m. tensors,
see (\ref{reconstruction}). What matters here is that the full one-loop e.m.
tensor contains the information about the e.m. tensor correlators with any
number of entries. The first non-trivial one corresponds of course to $n=2$.

Now let us apply the reconstruction formula (\ref{reconstruction}) to a single
chiral fermion theory. Classically
the energy-momentum tensor for a left-handed fermion is 
\be
T^{(L)}_{\mu\nu}= \frac i4 \overline {\psi_L} \gamma_\mu
{\stackrel{\leftrightarrow}{\del}}_\nu\psi_L+ \{\mu\leftrightarrow \nu\}
 \label{emt}
\ee
which is both conserved and traceless on shell. An analogous expression holds
for a right-handed fermion. It has been proved in general (and we have shown it
above) that the (unregulated) parity-odd $3$-point function in the coordinate
representation vanishes. Thus let us ask ourselves what would happen if
parity-odd amplitudes $$\langle 0|\mathcal{T}\{T^{(L)}_{\mu_1\nu_1}(x_1)\ldots
T^{(L)}_{\mu_n\nu_n}(x_n)\}|0\rangle_{\rm{odd}}$$ to all orders were to vanish.
We would have the same also for the right handed counterpart, while the
even-parity amplitudes are equal. Therefore the difference
 \be 
\langle\!\langle T^{(L)}_{\mu\nu}(x)\rangle\!\rangle- \langle\!\langle
T^{(R)}_{\mu\nu}(x)\rangle\!\rangle=0\label{TL-TR}
\ee
This would imply that the quantum analog of $\overline {\psi}
\gamma_\mu\gamma_5{\stackrel{\leftrightarrow}{\del}}_\nu\psi+
\{\mu\leftrightarrow \nu\}$ would vanish identically. This is nonsense, and
means that the vanishing of the parity-odd $3$-point function is an accidental
occurrence and that the (semiclassical) parity-odd amplitudes will generically
be non-vanishing \footnote{The analogue of the parity-odd $3$-point correlator
vanishing theorem does not exist for generic amplitudes.}. Inserting now these
results in the reconstruction formula (\ref{reconstruction})  and resuming the
series we would reconstruct the parity-odd anomaly. Let us apply this to the
trace of the quantum energy-momentum tensor. Since the parity-odd amplitudes are
generically nonvanishing we would obtain a nonvanishing trace anomaly. Now the
only possible covariant parity-odd anomaly is the Pontryagin density
\be
P=\frac
12\left(\epsilon^{nmlk}\ER_{nmpq}\ER_{lk}{}^{pq}\right)\label{pontryagin}
\ee
whose first nonvanishing contribution is quadratic in $h_{\mu\nu}$
\be
\epsilon^{\mu\nu\lambda \rho}R_{\mu\nu}{}^{\sigma\tau}R_{\lambda\rho\sigma\tau}=
2 \epsilon^{\mu\nu\lambda \rho} \left(\d_\mu\d_\sigma h^\tau_\nu \,
\d_\lambda\d_\tau h_{\rho}^\sigma-\d_\mu\d_\sigma h^\tau_\nu \,
\d_\lambda\d^\sigma h_{\tau\rho}\right)+\dots,\label{epsRR}
\ee
and can come only from the parity-odd $3$-point correlator. But, if the latter
vanishes, we would get an incomplete, and therefore non-covariant, expression
for this anomaly. 

The conclusion of this argument is: covariance (and consistency) requires that,
even if the (unregulated) parity-odd $3$-point function in the coordinate
representation vanishes, the corresponding regularized counterpart must be
non-vanishing. This is precisely what was found in \cite{BGL} with (regularized)
Feynman diagram techniques.  

The existence of the Pontryagin anomaly is confirmed also by other methods of calculation: the heat kernel method, see \cite{ChD2,BGL} and references therein, and the mass regularization of \cite{DS}, although the latter method have not been applied with the same accuracy as the dimensional regularization in the present paper. We should mention also the dispersive method which uses unitarity as an input. Of course we do not expect this method to reproduce this anomaly, which violates unitarity, \cite{BGL}. In fact using such a method would be a reversal of the burden of proof. The dispersive argument is very elegant and powerful, \cite{Cappelli:2001pz,Dolgov:1971ri,bert}, but it assume unitarity. Unitarity is normally given for granted and assumed by default. But the case presented in this paper is precisely an example in which this cannot be done.   

Finally we would like to notice that the so-called Delbourgo-Salam anomaly, \cite{DS}, i.e. the
anomaly in the divergence of the chiral current $j_{\mu 5}= i\bar \psi \gamma_\mu \gamma_5\psi$, 
is determined by a term (\ref{T3}) in which the factor $(2\slashed{p}-\slashed{q})$ is 
replaced by $\slashed{q}$. If, in such a term, we rewrite $ \slashed{q}$ as $2\slashed{p}-(2\slashed{p}-\slashed{q})$, 
we see that the second part reproduces the Pontryagin anomaly we have computed, while the term 
containing $2\slashed{p}$, once regularized, is easily seen to vanish. In other words
the Pontryagin trace anomaly and the Delbourgo-Salam chiral anomaly come from the same term.

\section{Conclusions}
 
In conclusion, let us summarize what was reviewed and what was shown in this paper. Our paradigm 
is always the theory of a free chiral fermion, thus every time that we refer to Feynman diagram 
techniques or Wick theorem, we are making reference to these techniques applied to this specific model. 

We started in sections \ref{sec:2ptFunctionIn2d}, \ref{sec:ParityOddIn2d} and \ref{sec:FeynmanDiagrams2d} 
by reviewing the regularization of the $2$-point function of e.m. tensors in $2d$, using both
differential regularization and dimensional regularization of the expression obtained with 
Feynman diagrams. Demanding the correlator to satisfy the Ward identity for diffeomorphism 
invariance we obtain a violation of the Ward identity for conformal invariance and we recover 
the known result of the $2d$ trace anomaly. In section \ref{sec:2ptFunction4d} the analogous 
result was shown also in $4d$ where the situation is different because we are able to regularize 
the correlator in such a way that both Ward identities are satisfied.

In section \ref{sec:3ptCorrelator}, moving to the $3$-point function of e.m. tensors in 
$4d$, we first noted a discrepancy between the computations in momentum space through 
Feynman diagrams and the computation in coordinate space using the Wick theorem. The direct c
omputation through Wick theorem tells us that there is no (unregulated) parity-odd 
contribution in the $3$-point correlator of e.m. tensors for the free chiral fermion. 
This result is indeed in agreement with the general fact that in $4d$ there are no 
parity-odd contribution in the correlation function of three e.m. tensors which was 
reviewed in section \ref{ssec:no-go}. With this fact in hand one could try to regularize 
this correlator with the techniques of differential regularization and would be obliged 
to conclude that there is no parity-odd trace anomaly simply because there is no 
parity-odd contribution to be regularized. On the other hand, by doing the 
computation in momentum space with Feynman diagram techniques we do find a 
parity-odd trace anomaly. Is this result forced to be wrong?

We argued in section \ref{sec:UglyDuckling} that these results can perfectly coexist and 
the result in coordinate space by no means is a no-go for the existence of the Pontryagin anomaly.

\acknowledgments

BLS would like to thank Alexander Zhiboedov for useful discussion. LB would like to thank Roberto Auzzi, 
Carl Bender, Maro Cvitan, Holger Nielsen, Silvio Pallua, Predrag Dominis-Prester, Ivica Smoli\'c, 
Alexander Sorin and in particular Adam Schwimmer 
for very interesting discussions. ADP acknowledges CNPq and CAPES for financial support and, in particular, 
CAPES for supporting his stay at SISSA and the Theoretical Particle Physics division of SISSA for 
supporting and hospitality. This work has been supported in part by
the Croatian Science Foundation under the project 8946.

\section*{Appendices}
\appendix

\section{Direct computation for a chiral fermion in $2d$}
\label{sec:2dchiralmodel}

Consider a free chiral fermion $\psi_{L}$ in $2d$ which has the $2$-point
function
\begin{equation}
\left\langle
\psi_{L}\left(x\right)\overline{\psi_{L}}\left(y\right)\right\rangle
=\frac{i}{2\pi}\frac{\gamma\cdot\left(x-y\right)}{\left(x-y\right)^{2}}P_{L},
\quad
P_{L}=\frac{1-\gamma_{*}}{2},\label{eq:PropagatorChiralFermion}
\end{equation}
and the e.m. tensor
\begin{equation}
T_{\mu\nu}=\frac{i}{4}\left(\overline{\psi_{L}}\gamma_{\mu}\stackrel{
\leftrightarrow}{\partial}_{\nu}\psi_{L}+\mu\leftrightarrow\nu\right).\label{
eq:EMTensorChiralFermion}
\end{equation}
Before proceeding with the calculation let us recall some definitions:
\begin{equation}
\left\{ \gamma^{\mu},\gamma^{\nu}\right\}
=2\eta^{\mu\nu}\Rightarrow\left(\gamma^{0}\right)^{2}=1,
\quad\left(\gamma^{i}\right)^{2}=-1.
\end{equation}
Clearly, $\gamma^{0}=\gamma_{0}$ and $\gamma^{i}=-\gamma_{i}$. For
an arbitrary dimension $D$ the analogous of $\gamma_{5}$ will be
denoted $\gamma_{*}$ and it is given by
$\gamma_{*}=\left(-i\right)^{\frac{D}{2}+1}\gamma_{0}\gamma_{1}\dots\gamma_{D-1}
,$
which for $D=2$ means $\gamma_{*}=-\gamma_{0}\gamma_{1}.$
 
It is straightforward to check that the following relations are true:
\begin{equation}
\gamma_{\mu}=\epsilon_{\mu\nu}\gamma^{\nu}\gamma_{*},\quad\epsilon_{\mu\nu}
\gamma^{\nu}=\gamma_{\mu}\gamma_{*},\label{eq:UsefulIdentity}
\end{equation}
where we are using the convention where $\epsilon_{01}=1$. It follows
\begin{equation}
\trace{\left(\gamma_{\mu}\gamma_{\nu}\gamma_{*}\right)}=-2\epsilon_{\mu\nu}
.\label{eq:TraceWithGammaStar}
\end{equation}

Our purpose is to compute the $2$-point of the em tensor in the theory
(\ref{eq:PropagatorChiralFermion}).
Since we are dealing with a simple free theory we can use the Wick theorem.

The non-zero part of the correlation function comes from the 
\[
\left\langle T_{\mu\nu}\left(x\right)T_{\rho\sigma}\left(y\right)\right\rangle
=\frac{1}{16}\left\langle
:\bar{\psi}_{L}\gamma_{\mu}\stackrel{\leftrightarrow}{\partial}_{\nu}\psi_{L}
:\left(x\right):\bar{\psi}_{L}\gamma_{\rho}\stackrel{\leftrightarrow}{\partial}_
{\sigma}\psi_{L}:\left(y\right)\right\rangle +\text{sym.},
\]
which is given by the full contraction of this object, namely
\be \begin{aligned}
\left\langle T_{\mu\nu}\left(x\right)T_{\rho\sigma}\left(y\right)\right\rangle
=&
\frac{1}{16(2\pi)^2}\left(\trace{\left[\gamma_{\mu}\partial_{\nu}^{x}
\left\langle
\psi_{L}\left(x\right)\bar{\psi}_{L}\left(y\right)\right\rangle
\gamma_{\rho}\partial_{\sigma}^{y}\left\langle
\psi_{L}\left(y\right)\bar{\psi}_{L}\left(x\right)\right\rangle
\right]} + \cdots\right)\\
&+\rm{sym. },
\end{aligned}\ee
where the ellipsis stand for the three other ways of organizing the derivatives.
We may use the translational invariance of this correlator to shift
$x\rightarrow x-y$ and $y\rightarrow0$. For simplicity we will relabel
$x-y$ calling it simply $x$. Since the correlation function is simply
a function of $x-y$, $\partial^{y}=-\partial^{x}$. Let us also remark
that $\left\langle
\psi_{L}\left(x\right)\bar{\psi}_{L}\left(y\right)\right\rangle =-\left\langle
\psi_{L}\left(y\right)\bar{\psi}_{L}\left(x\right)\right\rangle $.
Thus, we can exchange all the derivatives on $y$ by derivatives on
$x$ and the correlations functions $\left\langle
\psi_{L}\left(y\right)\bar{\psi}_{L}\left(x\right)\right\rangle $
by $\left\langle \psi_{L}\left(x\right)\bar{\psi}_{L}\left(y\right)\right\rangle
$,
which, due to translational invariance, can be written as $\left\langle
\psi_{L}\left(x-y\right)\bar{\psi}_{L}\left(0\right)\right\rangle $.
Therefore,
\be \begin{aligned}
\left\langle T_{\mu\nu}\left(x\right)T_{\rho\sigma}\left(y\right)\right\rangle =
& \frac{1}{16(2\pi)^2} \left(
\trace{\left[\gamma_{\mu}\partial_{\nu}\left\langle
\psi_{L}\left(x\right)\bar{\psi}_{L}\left(0\right)\right\rangle
\gamma_{\rho}\partial_{\sigma}\left\langle
\psi_{L}\left(x\right)\bar{\psi}_{L}\left(0\right)\right\rangle
\right]} + \cdots \right)\\
& +\rm{sym.}
\end{aligned}\ee
Using the expression for the $2$-point function
\eqref{eq:PropagatorChiralFermion} we have
\be
\trace{\left[\gamma_{\mu}\partial_{\nu}\left\langle
\psi_{L}\left(x\right)\bar{\psi}_{L}\left(0\right)\right\rangle
\gamma_{\rho}\partial_{\sigma}\left\langle
\psi_{L}\left(x\right)\bar{\psi}_{L}\left(0\right)\right\rangle \right]} & = &
\frac{1}{(2\pi)^2}\partial_{\nu}\left(\frac{x^{\alpha}}{x^{2}}\right)\partial_{
\sigma}\left(\frac{
x^{\beta}}{x^{2}}\right)\trace{\left(\gamma_{\mu}\gamma_{\alpha}\gamma_{\rho}
\gamma_{\beta}P_{L}\right)}\0,
\ee
and analogously for the other terms. One should notice that
$$\trace{\left(\gamma_{\mu}\gamma_{\beta}\gamma_{\rho}\gamma_{\alpha}P_{L}
\right)
}=\trace{\left(\gamma_{\rho}\gamma_{\alpha}\gamma_{\mu}\gamma_{\beta}P_{L}
\right)}$$
and we are able to rewrite our correlation function as
\begin{multline}
\left\langle T_{\mu\nu}\left(x\right)T_{\rho\sigma}\left(y\right)\right\rangle
=
\frac{1}{16}\frac{1}{(2\pi)^2}\left[\partial_{\nu}\left(\frac{x^{\alpha}}{x^{2}}
\right)\partial_{
\sigma}\left(\frac{x^{\beta}}{x^{2}}\right)-\left(\frac{x^{\alpha}}{x^{2}}
\right)\partial_{\nu}\partial_{\sigma}\left(\frac{x^{\beta}}{x^{2}}\right)\right
]\\
\times\left[\trace{\left(\gamma_{\mu}\gamma_{\alpha}\gamma_{\rho}\gamma_{\beta}
P_{L}
\right)}+\mu\leftrightarrow\rho\right]+\rm{sym}.\label{eq:NeedToEval}
\end{multline}
Exchanging the position of $\gamma_{\alpha}$ and $\gamma_{\rho}$
in
$\trace{\left(\gamma_{\mu}\gamma_{\alpha}\gamma_{\rho}\gamma_{\beta}P_{L}\right)
}$
we have
\begin{equation}
\trace{\left(\gamma_{\mu}\gamma_{\alpha}\gamma_{\rho}\gamma_{\beta}P_{L}\right)}
=2\eta_{\alpha\rho}\trace{\left(\gamma_{\mu}\gamma_{\beta}P_{L}\right)}-\trace{
\left(\gamma_{\mu}\gamma_{\rho}\gamma_{\alpha}\gamma_{\beta}P_{L}\right)}.\0
\end{equation}
Thus
\begin{eqnarray}
\trace{\left(\gamma_{\mu}\gamma_{\alpha}\gamma_{\rho}\gamma_{\beta}P_{L}\right)}
+\mu\leftrightarrow\rho & = &
2\eta_{\alpha\rho}\trace{\left(\gamma_{\mu}\gamma_{\beta}P_{L}\right)}+2\eta_{
\alpha\mu}\trace{\left(\gamma_{\rho}\gamma_{\beta}P_{L}\right)}-\trace{
\left(\left\{ \gamma_{\mu},\gamma_{\rho}\right\}
\gamma_{\alpha}\gamma_{\beta}P_{L}\right)}\nonumber \0\\
 & = &
2\left[\eta_{\alpha\rho}\trace{\left(\gamma_{\mu}\gamma_{\beta}P_{L}\right)}
+\eta_{\alpha\mu}\trace{\left(\gamma_{\rho}\gamma_{\beta}P_{L}\right)}-\eta_{
\mu\rho}\trace{\left(\gamma_{\alpha}\gamma_{\beta}P_{L}\right)}\right].\0
\end{eqnarray}
The trace of $\gamma_{\mu}\gamma_{\nu}P_{L}$ is straightforward to compute (see
Appendix):
\begin{equation}
\trace{\left(\gamma_{\mu}\gamma_{\nu}P_{L}\right)}=\frac{1}{2}\left[\trace{
\left(\gamma_{\mu}\gamma_{\nu}\right)}-\trace{\left(\gamma_{\mu}\gamma_{\nu}
\gamma_{*}\right)}\right]=\eta_{\mu\nu}+\epsilon_{\mu\nu}.\0
\end{equation}
Therefore
\begin{equation}
\trace{\left(\gamma_{\mu}\gamma_{\alpha}\gamma_{\rho}\gamma_{\beta}P_{L}\right)}
+\mu\leftrightarrow\rho=2\left(\eta_{\alpha\rho}\eta_{\mu\beta}+\eta_{\alpha\mu}
\eta_{\rho\beta}-\eta_{\mu\rho}\eta_{\alpha\beta}\right)+2\left(\eta_{\alpha\rho
}\epsilon_{\mu\beta}+\eta_{\alpha\mu}\epsilon_{\rho\beta}-\eta_{\mu\rho}
\epsilon_{\alpha\beta}\right).\label{eq:FirstForm}
\end{equation}
It turns out that we are able to rewrite $\eta_{\mu\rho}\epsilon_{\alpha\beta}$
as
\[
\eta_{\mu\rho}\epsilon_{\alpha\beta}=\frac{1}{2}\left(\eta_{\alpha\mu}\epsilon_{
\rho\beta}-\eta_{\beta\mu}\epsilon_{\rho\alpha}+\eta_{\alpha\rho}\epsilon_{
\mu\beta}-\eta_{\beta\rho}\epsilon_{\mu\alpha}\right)\0
\]
and using this expression we may rewrite \eqref{eq:FirstForm} as
\begin{eqnarray}
\trace{\left(\gamma_{\mu}\gamma_{\alpha}\gamma_{\rho}\gamma_{\beta}P_{L}\right)}
+\mu\leftrightarrow\rho & = &
2\left(\eta_{\alpha\rho}\eta_{\mu\beta}+\eta_{\alpha\mu}\eta_{\rho\beta}-\eta_{
\mu\rho}\eta_{\alpha\beta}\right)\nonumber \\
 &  &
+\left(\eta_{\alpha\rho}\epsilon_{\mu\beta}+\eta_{\alpha\mu}\epsilon_{\rho\beta}
+\eta_{\beta\mu}\epsilon_{\rho\alpha}+\eta_{\beta\rho}\epsilon_{\mu\alpha}
\right).\label{eq:SecondForm}
\end{eqnarray}
Using \eqref{eq:SecondForm} we can compute \eqref{eq:NeedToEval}
and we find the parity-odd part
\be \begin{aligned}
&\left\langle T_{\mu\nu}\left(x\right)T_{\rho\sigma}\left(0\right)\right\rangle
_{\text{odd}} = \\
 & -\frac{1}{4\pi^2} \left(
\frac{\epsilon_{\mu\alpha}x^{\alpha}x_{\nu}x_{\rho}x_{\sigma}}{x^{8}}+\frac{
\epsilon_{\nu\alpha}x^{\alpha}x_{\mu}x_{\rho}x_{\sigma}}{x^{8}}+\frac{\epsilon_{
\rho\alpha}x^{\alpha}x_{\mu}x_{\nu}x_{\sigma}}{x^{8}}+\frac{\epsilon_{
\sigma\alpha}x^{\alpha}x_{\mu}x_{\nu}x_{\rho}}{x^{8}}\right. \\
 &  
\phantom{-\frac{1}{4\pi^2}}-\frac{\epsilon_{\mu\alpha}\eta_{\rho\nu}x^{\alpha}x_
{\sigma}}{4x^{6}}-\frac{
\epsilon_{\mu\alpha}\eta_{\sigma\nu}x^{\alpha}x_{\rho}}{4x^{6}}-\frac{\epsilon_{
\nu\alpha}\eta_{\rho\mu}x^{\alpha}x_{\sigma}}{4x^{6}}-\frac{\epsilon_{\nu\alpha}
\eta_{\sigma\mu}x^{\alpha}x_{\rho}}{4x^{6}} \\
 &  
\phantom{-\frac{1}{4\pi^2}}\left.-\frac{\epsilon_{\rho\alpha}\eta_{\mu\sigma}x^{
\alpha}x_{\nu}}{4x^{6}}-\frac{
\epsilon_{\rho\alpha}\eta_{\nu\sigma}x^{\alpha}x_{\mu}}{4x^{6}}-\frac{\epsilon_{
\sigma\alpha}\eta_{\mu\rho}x^{\alpha}x_{\nu}}{4x^{6}}-\frac{\epsilon_{
\sigma\alpha}\eta_{\nu\rho}x^{\alpha}x_{\mu}}{4x^{6}}\right).\label{eq:ResultChiralFermion2d}
\end{aligned}\ee
As a matter of fact, out of this computation we find that the parity-even
part matches \eqref{eq:Goal} with $c=1/4\pi^2$, in agreement with
\cite{Osborn93,Osborn96}. The expression
\eqref{eq:ResultChiralFermion2d}
is traceless, conserved and can be written as
\begin{equation}
\left\langle T_{\mu\nu}\left(x\right)T_{\rho\sigma}\left(0\right)\right\rangle
_{\text{odd}}=\frac{1}{32\pi^2}\left(\epsilon_{\alpha\mu}T_{\phantom{\alpha}
\nu\rho\sigma}^{\alpha}+\epsilon_{\alpha\nu}T_{\mu\phantom{\alpha}\rho\sigma}^{
\phantom{\mu}\alpha}+\epsilon_{\alpha\rho}T_{\mu\nu\phantom{\alpha}\sigma}^{
\phantom{\mu\nu}\alpha}+\epsilon_{\alpha\rho}T_{\mu\nu\rho\phantom{\alpha}}^{
\phantom{\mu\nu\rho}\alpha}\right),\label{eq:ExampleWithParityEven}
\end{equation}
where $T_{\mu\nu\rho\sigma}$ is given by the expression (\ref{Smnrs}). Hence
\eqref{eq:ResultChiralFermion2d} agrees with the null cone
result.

\section{The chiral fermion model in $4d$}
\label{sec:chiralm}

In this appendix we summarize the formalism and notation of \cite{BGL},
concerning the free chiral fermion model minimally coupled to gravity. The
action is
\be 
S= \int d^4x \, \sqrt{|g|} \,\left[ \frac i2\overline {\psi_R} \gamma^\mu
{\stackrel{\leftrightarrow}{\d}}_\mu  \psi_R -\frac 14\epsilon^{\mu a b c}
\omega_{\mu a b} \overline{\psi_R} \gamma_c \gamma_5\psi_R\right]
\label{action1}
\ee
where it is understood that the derivative applies to $\psi_L$ and $\overline
{\psi_L}$ only. We have used the relation $\{\gamma^a, \Sigma^{bc}\}=i \,
\epsilon^{abcd}\gamma_d\gamma_5$.
Now one expands
\be
e_\mu^a = \delta_\mu^a +\chi_\mu^a+\dots,\quad\quad e_a^\mu = \delta _a^\mu
+\hat
\chi_a^\mu +\dots,\quad\quad {\rm and}\quad
g_{\mu\nu}=\eta_{\mu\nu}+h_{\mu\nu}+\dots \, .\label{hmunu}
\ee
Using the defining relations of metric and vierbein one finds
\be
\hat \chi_\nu^\mu =- \chi_\nu^\mu\quad \quad {\rm and}\quad\quad
h_{\mu\nu}=2\,\chi_{\mu\nu}.\label{hatchichi}
\ee
The spin connection $\omega_m= \omega_m^{ab} \Sigma_{ab}$,
where $\Sigma_{ab} = \frac 14 [\gamma_a,\gamma_b]$ are the Lorentz generators,
to lowest order is
\be 
\omega_{\mu a b}\, \epsilon^{\mu a b c}= - \epsilon^{\mu a b c}\,  \d_\mu
\chi_{a\lambda}\,\chi_b^\lambda+\dots\, .\label{omega}
\ee
Therefore up to second order the action can be written
\be 
S = \int d^4x \,  \left[\frac i2 (\delta^\mu_a -\chi^\mu_a ) \overline
{\psi_L} \gamma^a {\stackrel{\leftrightarrow}{\d}}_\mu  \psi_L +\frac
14\epsilon^{\mu a b c}\,  \d_\mu \chi_{a\lambda}\,\chi_b^\lambda\, 
\overline\psi_L \gamma_c \gamma_5\psi_L\right].
\ee
Splitting it into free and interacting parts, one can extract Feynman rules. The
fermion propagator is
\be 
\frac i{\slash \!\!\! p+i\epsilon }\label{prop}
\ee
The two-fermion-one-graviton vertex ($V_{ffg}$) 
\be
-\frac i{8} \left[(p-p')_\mu \gamma_\nu + (p-p')_\nu \gamma_\mu\right] \frac
{1+\gamma_5}2\label{2f1g}
\ee
The two-fermion-two-graviton vertex ($V_{ffgg}$) is
\be 
\frac 1{64} t_{\mu\nu\mu'\nu'\kappa\lambda}(k-k')^\lambda\gamma^\kappa\frac
{1+\gamma_5}2\label{2f2g}
\ee 
where
\be
t_{\mu\nu\mu'\nu'\kappa\lambda}=\eta_{\mu\mu'} \epsilon_{\nu\nu'\kappa\lambda}
+\eta_{\nu\nu'} \epsilon_{\mu\mu'\kappa\lambda} +\eta_{\mu\nu'}
\epsilon_{\nu\mu'\kappa\lambda} +\eta_{\nu\mu'}
\epsilon_{\mu\nu'\kappa\lambda}\label{t}
\ee

\section{Regularization formulas in $2d$ and $4d$}
\label{sec:regformulas}

In this appendix we collect the regularized integrals that are needed to
evaluate the Feynman diagrams in the text both in $2d$ and $4d$. The integrals
below are {\it Euclidean integrals}. They are an intermediate results needed in
order to compute the Feynman diagrams in the text. Since the starting points and
the final results are Lorentzian, it is understood that one has to do the
appropriate Wick rotations in order to be able to use them.

In $2d$, after introducing $\delta$ extra dimensions in the internal momentum
and a Feynman parameter $u$ ($0\leq u\leq1$), in the limit $\delta\to 0$, we
have  
\be
&&  \int\!\frac{d^2p}{(2\pi)^2}\int\!\frac{d^\delta \ell}{(2\pi)^\delta}\frac
{\ell^2}{({p}^2+\ell^2+\Delta)^2}=-\frac 1{4\pi}\0\\
&&  \int\!\frac{d^2p}{(2\pi)^2}\int\!\frac{d^\delta \ell}{(2\pi)^\delta}\frac
{\ell^2 p^2}{({p}^2+\ell^2+\Delta)^2}=\frac 1{4\pi}\Delta\label{C2b}
\ee
and 
\be
&&   \int\!\frac{d^2p}{(2\pi)^2}\int\!\frac{d^\delta \ell}{(2\pi)^\delta}\frac
{p^2}{({p}^2+\ell^2+\Delta)^2}=
\frac 1{4\pi}\frac 1{\Delta}\0\\
&&   \int\!\frac{d^2p}{(2\pi)^2}\int\!\frac{d^\delta \ell}{(2\pi)^\delta}\frac
{p^2}{({p}^2+\ell^2+\Delta)^2}=
\frac 1{4\pi}\left(- \frac 2{\delta}-\gamma +\log (4\pi)-
\log{\Delta}\right)\0\\
&&  \int\!\frac{d^2p}{(2\pi)^2}\int\!\frac{d^\delta \ell}{(2\pi)^\delta}\frac
{p^4}{({p}^2+\ell^2+\Delta)^2}=
\frac 1{2\pi}\Delta\left( \frac 2{\delta}-1+\gamma -\log
(4\pi)+\log{\Delta}\right)\label{C2a}
\ee
where $\Delta=u(1-u)k^2$.

Proceeding in the same way in $4d$, with two Feynman parameters $u$ and $v$, in
the limit $\delta\to 0$, beside (\ref{basic}), we find 
\be
&&   \int\!\frac{d^4p}{(2\pi)^4}\int\!\frac{d^\delta \ell}{(2\pi)^\delta}\frac
{p^2}{({p}^2+\ell^2+\Delta)^3}=
\frac 1{(4\pi)^2}\left(- \frac 2{\delta}-\gamma +\log (4\pi)-
\log{\Delta}\right)\0\\
&&  \int\!\frac{d^4p}{(2\pi)^4}\int\!\frac{d^\delta \ell}{(2\pi)^\delta}\frac
{p^4}{({p}^2+\ell^2+\Delta)^3}=
\frac {\Delta}{2(4\pi)^2}\left(- \frac 2{\delta}-\gamma +4+\log
(4\pi)-\log{\Delta}\right)\label{C4a}
\ee
and
\be
&&  \int\!\frac{d^4p}{(2\pi)^4}\int\!\frac{d^\delta \ell}{(2\pi)^\delta}\frac
{\ell^2}{({p}^2+\ell^2+\Delta)^3}=-\frac 1{2(4\pi)^2}\0\\
&&  \int\!\frac{d^4p}{(2\pi)^4}\int\!\frac{d^\delta \ell}{(2\pi)^\delta}\frac
{\ell^2 p^2}{({p}^2+\ell^2+\Delta)^3}=
\frac 1{(4\pi)^2}\Delta\label{C4b}
\ee
where $\Delta = u(1-u) k_1+v(1-v) k_2 + 2 uv \,k_1k_2$.

\section{Fourier transforms}
\label{sec:Fourier}

In this appendix we expand on the results of section (\ref{ssec:Fourier}). Let
us start from the following formal transformations:
\be \begin{aligned}
&\!\!\!\!\!\!\!\! -i(2\pi)^6 \int\! \frac{d^4p}{(2\pi)^4}\, \frac
{k_{1\mu}}{p^2(p-k_1)^2(p-q)^2}=i\int\! d^4x\, d^4y\, e^{i(k_1 x+k_2 y)} \, 
\frac {\del}{\del x^\mu}\left( \frac 1{(x-y)^2 x^2 y^2}\right)\\
&\!\!\!\!\!\!\!\! -i(2\pi)^6\int\! \frac{d^4p}{(2\pi)^4}\, \frac
{k_{2\mu}}{p^2(p-k_1)^2(p-q)^2}=i\int\! d^4x\, d^4y\, e^{i(k_1 x+k_2 y)} \, 
\frac {\del}{\del y^\mu}\left(\frac 1{(x-y)^2 x^2 y^2}\right) \label{C1}
\end{aligned}\ee
According to the procedure outlined in section (\ref{ssec:Fourier}), the LHS's 
of these equations will be defined by means of (\ref{basicreg}) and, via Fourier 
anti-transform, will define the corresponding regularized rational function in the RHS's.
The generalization to multiple powers of the momenta $k_1,k_2$ in the numerator
is straightforward. The (\ref{C1}) formulas and the like define a {\it differential regularization}.

In the main body of the paper we have to do with similar integrals in which, however,
the numerator of the integrand contains polynomials of $p$ beside $k_1$ and $k_2$.  
In this case we do not know a straightforward way to differentially regularize them and
resort instead to {\it dimensional regularization}, in which case
other Fourier transforms are needed. For instance
\be 
&&\int \frac{d^4k_1}{(2\pi)^4}  \frac{d^4k_2}{(2\pi)^4} \frac {e^{i (k_1(x-z)
+k_2(y-z))}}{(k_1+k_2)^2} \label{C3}\\
&&= \frac 1{16}\left(\int \frac{d^4\tilde k_1}{(2\pi)^4}\frac {e^{i \tilde
k_1\left(\frac{(x-z) +(y-z)}2\right)}}{\tilde k_1^2}\right)
\left(\int \frac{d^4\tilde k_2}{(2\pi)^4}\, e^{i\tilde k_2\left(\frac
{x-y}2\right)}\right)=\frac 1{16\pi^2} \frac 1{(x-z)^2} \delta^{(4)}(x-y)\0
\ee
where we set $\tilde k_1=k_1+k_2$ and $\tilde k_2=k_1-k_2$. Proceeding in the same way,
\begin{multline}
\int \frac{d^4k_1}{(2\pi)^4}  \frac{d^4k_2}{(2\pi)^4} \frac {e^{i (k_1(x-z)
+k_2(y-z))}}{(k_1+k_2)^2}\log{(k_1+k_2)^2}= \\
= \frac 1{4\pi^2}  \delta^{(4)}(x-y) \frac 1{(x-z)^2}\log
\frac{(x-z)^2}4,\label{C4}
\end{multline}
and it is understood that
\begin{multline*}
\int \frac{d^4k_1}{(2\pi)^4}  \frac{d^4k_2}{(2\pi)^4} \, {e^{i (k_1(x-z)
+k_2(y-z)}}\log{(k_1+k_2)^2}=\\
= -(\del_x+\del_y)^2\int \frac{d^4k_1}{(2\pi)^4}  \frac{d^4k_2}{(2\pi)^4} 
\frac {e^{i (k_1(x-z) +k_2(y-z)}}{(k_1+k_2)^2}\log{(k_1+k_2)^2}.
\end{multline*}

\section{Conservation of the e.m. tensor}
\label{sec:divergence}

In this appendix we complete the proof of section \ref{ssec:divergence}.

To start with we write down the structure of the various terms in (\ref{Div2})
in momentum representation and in 
coordinate space after applying (\ref{reconstruction})
\be
\epsilon_{\nu\beta\rho\tau}k_1\!\cdot \! k_2 k_{1\la}
k_{1\al}k_1^\tau\rightarrow &&
\int \xi^\nu\epsilon_{\nu\beta\rho\tau} \del_\sigma \del_\la\del_\al \del^\tau
h^{\la\rho} \del^\sigma h^{\al\beta}=0,\0\\
\epsilon_{\nu\beta\rho\tau}k_1\!\cdot \! k_2 k_{1\la}
k_{1\al}k_2^\tau\rightarrow 
&&\int \xi^\nu\epsilon_{\nu\beta\rho\tau} \del_\sigma \del_\la\del_\al 
h^{\la\rho}\, \del^\tau \del^\sigma h^{\al\beta}\label{Div3}\\
&&=\frac 12\int \xi^\nu\epsilon_{\nu\beta\rho\tau} \del_\sigma \del^\rho\del_\al
h\, \del^\tau \del^\sigma h^{\al\beta},\\
\epsilon_{\nu\beta\rho\tau}k_1\!\cdot \! k_2 k_{1\la}
k_{2\al}k_2^\tau\rightarrow 
&&\int \xi^\nu\epsilon_{\nu\beta\rho\tau} \del_\sigma \del_\al h^{\la\rho}\,
\del_\la\del^\tau \del^\sigma h^{\al\beta},\label{Div4}\\
\epsilon_{\nu\beta\rho\tau}k_1\!\cdot \! k_2 k_{1\la}
k_{2\al}k_1^\tau\rightarrow 
&&\int \xi^\nu\epsilon_{\nu\beta\rho\tau} \del_\sigma \del_\al \del^\tau
h^{\la\rho}\, \del_\la\del^\sigma h^{\al\beta},\label{Div5}
\ee
and other similar terms obtained by exchanging 1 and 2. (\ref{Div4}) is the
opposite of (\ref{Div5}). In addition we have the term
\be 
\eta_{\alpha\la}\epsilon_{\nu\beta\rho\tau}(k_1\!\cdot \! k_2)^2 
k_1^\tau\rightarrow 
&&\int \xi^\nu\epsilon_{\nu\beta\rho\tau} \del_\sigma\del_\kappa \del^\tau
h^{\la\rho} \del^\kappa\del^\sigma h^{\al\beta}\label{Div6}
\ee
and the opposite one obtained by exchanging 1 and 2. All these terms appear with
(nonvanishing) coefficients which are rational numbers or 
rational numbers multiplied by 
\be 
\frac 2{\delta} +\gamma- \log 4\pi + \log 2k_1\!\cdot\!k_2\label{infinitefactor}
\ee
in the limit $\delta\to 0$. The terms proportional to $\log 2k_1\!\cdot\!k_2$ will 
be disregarded here because, due to the results in appendix \ref{sec:Fourier}, they 
corresponds to the $2$-point terms of eq.(\ref{2nddivergence}). All the other terms 
have to be canceled by subtracting counterterms from the action. The important point 
is that such
counterterm must be Weyl invariant to the appropriate order in $h$, otherwise they 
would modify the trace of the e.m. tensor.  We show next that this is in fact true for all the above terms. 

The terms (\ref{Div3}) and (\ref{Div6}) are trivial, for we have
\be 
&&\delta_\omega \int h\, \epsilon_{\mu\nu\la\rho}\del^\mu h^{\tau \nu}\,\del^\la
\square h^\rho_\tau=0,\0\\
&&\delta_\xi \int h\, \epsilon_{\mu\nu\la\rho}\del^\mu h^{\tau \nu}\,\del^\la
\square h^\rho_\tau=\int\xi^\nu \epsilon_{\nu\si\rho\la}
\del_\tau \del^\la\del^\kappa h \, \del^\si\del_\kappa
h^{\rho\tau},\label{trivial1}
\ee
and
\be 
&&\delta_\omega \int \epsilon_{\mu\nu\la\rho}\, h^{\mu\si}\, \del^\tau\del^\la 
h^\rho_\si\,\square h^\nu_\tau=0,\0\\
&&\delta_\xi  \int\epsilon_{\mu\nu\la\rho}\, h^{\mu\si}\, \del^\tau\del^\la 
h^\rho_\si\,\square h^\nu_\tau=-\int\xi^\nu \epsilon_{\nu\tau\la\rho}
\del^\tau\del^\kappa h^{\mu\si} \, \del_\mu\del_\kappa \del^\la
h^\rho_\si, \label{trivial2}\\
&&\quad+2\int \xi^\nu \epsilon_{\nu\tau\la\rho} \del^\kappa\del^\al
h^{\tau\si}\del_\kappa\del_\al \del^\la h^\rho_\si+
\frac 12 \int \xi^\nu \epsilon_{\nu\mu\tau\la} \del^\tau\del^\kappa
h^{\mu\si}\del_\kappa \del^\la \del_\si h.\0
\ee
Similarly
\be 
 \delta_\omega \int\, \epsilon_{\nu\beta\rho\tau}\, h^{\nu\al}\del_\kappa h^{\si
\rho}\,\del_\si \del^\tau\del^\kappa h^\beta_\al&=&0,\0\\
 \delta_\xi \int \epsilon_{\nu\beta\rho\tau}\, h^{\nu\al}\del_\kappa h^{\si
\rho}\,\del_\si \del^\tau\del^\kappa h^\beta_\al&=&
\int \xi^\nu \epsilon_{\nu\rho\beta\tau}\, \del_\si\del_\kappa h^{\al \rho}
\,\del^\si\del^\kappa \del^\tau h_\al^\beta\0\\
&&-\frac 12 \int \xi^\nu \epsilon_{\nu\rho\beta\tau}\, \del^\rho\del_\kappa
h^{\si \tau} \,\del_\si\del^\kappa \del^\beta h,\label{trivial3}
\ee
and 
\be 
&&\delta_\omega \int \epsilon_{\nu\beta\rho\tau}\, h^{\nu \si}\del_\kappa 
h^{\al\rho}\, \del^\tau\del_\si\del^\kappa h_\al^\beta=0,\0\\
&&\delta_\xi \int \epsilon_{\nu\beta\rho\tau}\, h^{\nu \si}\del_\kappa 
h^{\al\rho}\, \del^\tau\del_\si\del^\kappa h_\al^\beta  
=-\int\, \xi^\nu\Big{(}2\epsilon_{\nu\beta\rho\tau}\, \del_\kappa\del^\la
h^{\al\rho}\, \del_\al\del^\kappa \del^\tau h^\beta_\al
\label{trivial4}\\
&&+2  \epsilon_{\nu\rho\beta\tau}\, \del_\kappa\del^\rho h^{\al\si}\,
\del_\si\del^\kappa \del^\tau h_\al^\beta +2  \epsilon_{\nu\beta\rho\tau}\,
\del^\kappa\del^\tau \del_\si h_\al^\beta\, \del^\al \del_\kappa h^{\rho\si}
+ \epsilon_{\nu\beta\rho\tau}\,\del_\kappa\del^\rho h^{\beta\si}\, \del^\kappa
\del^\tau\del_\si h\Big{)},\0
\ee
as well as 
\be 
&&\delta_\omega \int \epsilon_{\nu\beta\rho\tau}\, h^{\nu \si}\square 
h^{\al\rho}\, \del^\tau\del_\si h_\al^\beta=0,\0\\
&&\delta_\xi \int \epsilon_{\nu\beta\rho\tau}\, h^{\nu \si}\square 
h^{\al\rho}\, \del^\tau\del_\si h_\al^\beta   
=\int\, \xi^\nu\Big{(}\frac 12\epsilon_{\nu\beta\rho\tau}\, \del_\kappa\del^\rho
h^{\beta\si}\, \del_\si\del^\kappa \del^\tau h
\label{trivial5}\\
&&\quad\quad -2  \epsilon_{\nu\rho\beta\tau}\, \del_\kappa\del^\al h^{\rho\si}\,
\del_\si\del^\kappa \del^\tau h_\al^\beta 
-  \epsilon_{\nu\beta\rho\tau}\,\del_\kappa\del^\rho h^{\al\si}\, \del^\kappa
\del^\tau\del_\si h_\al^\beta\Big{)},\0
\ee
and other similar ones. Using combinations of these relations it is easy to see 
that all the terms listed above, which appear in (\ref{Div2}), see (\ref{Div3}), 
(\ref{Div4}) and (\ref{Div5}), are in fact trivial. They can be reabsorbed in a 
redefinition of the action without altering the already calculated trace anomaly.

\end{document}